%% file: main.tex
\documentclass[iop, twocolappendix]{emulateapj}
\usepackage{amsmath}

\usepackage{comment}

\input{preamble}
\slugcomment{Accepted By ApJ}

\makeatletter
\renewcommand\normalsize{\@setfontsize\normalsize\@xpt{12.5}}
\makeatother

\usepackage{xcolor}
\definecolor{dlcolor}{HTML}{FF7F00}

\graphicspath{{Fig/}}        

\shorttitle{\Lcii--\,SFR and \cii Luminosity Function of \ncode{Simba} Galaxies at the EoR}
\shortauthors{Leung et al.}

\begin{document}
\title{Predictions of the \Lcii--\,SFR and \cii Luminosity Function at the Epoch of Reionization}

\author{T. K. Daisy Leung\altaffilmark{1}}
\author{Karen P. Olsen\altaffilmark{2}}
\author{Rachel S. Somerville\altaffilmark{1,3}}
\author{Romeel Dav\'e\altaffilmark{4}}
\author{Thomas R. Greve\altaffilmark{5, 6}}
\author{Christopher C. Hayward\altaffilmark{1}}
\author{Desika Narayanan\altaffilmark{7, 8, 9}}
\author{Gerg\"o Popping\altaffilmark{10}}

\affil{\textsuperscript{1} Center for Computational Astrophysics, Flatiron Institute, 162 Fifth Avenue, New York, NY 10010, USA; }
\email{dleung@flatironinstitute.org}
\altaffiltext{2}{Department of Astronomy and Steward Observatory, University of Arizona, Tucson, AZ 85721, USA}
\altaffiltext{3}{Department of Physics and Astronomy, Rutgers University, 136 Frelinghuysen Road, Piscataway, NJ 08854, USA}
\altaffiltext{4}{Institute for Astronomy, Royal Observatory, Univ. of Edinburgh, Edinburgh EH9 3HJ, UK}
\altaffiltext{5}{Cosmic Dawn Center (DAWN), DTU-Space, Technical University of Denmark, Elektrovej 327, DK-2800 Kgs.}
\altaffiltext{6}{Department of Physics and Astronomy, University College London, Gower Street, London WC1E 6BT, UK}
\altaffiltext{7}{Department of Astronomy, University of Florida, 211 Bryant Space Sciences Center, Gainesville, FL 32611 USA}
\altaffiltext{8}{University of Florida Informatics Institute, 432 Newell Drive, CISE Bldg E251, Gainesville, FL 32611}
\altaffiltext{9}{Cosmic Dawn Center at the Niels Bohr Institute, University of Copenhagen and DTU-Space, Technical University of Denmark}
\altaffiltext{10}{European Southern Observatory, Karl-Schwarzschild-Straße 2, D-85748 Garching, Germany}

 \begin{abstract}
We present the first predictions for the \Lcii--\,SFR relation and \cii luminosity function (LF) in the
Epoch of Reionization (EoR) based on
cosmological hydrodynamics simulations using the \simba suite plus radiative transfer calculations via \sigame.
The sample consists of 11,137 galaxies covering
halo mass $\log$~M$_{\rm halo}\in[9, 12.4]$\,\Msun, star formation rate SFR$\in[0.01, 330]$\,\Msun\,yr\pmOne, and metallicity
$\langle Z_{\rm gas}\rangle_{\rm SFR}\in[0.1,1.9]\,Z_\odot$. The simulated \Lcii-SFR relation is consistent with the range observed, but with a spread of $\simeq$\,0.3\,dex at the high end of SFR ($>$\,100\,\Msun\,yr\pmOne)
and $\simeq$0.6\,dex at the lower end, and there is tension between our predictions and the
values of \Lcii above 10$^{8.5}$\,\Lsun observed in some galaxies reported in the literature.
The scatter in the \Lcii--\,SFR relation is mostly driven by galaxy properties, 
such that at a given SFR, galaxies with higher molecular mass and metallicity have higher \Lcii.
The \cii LF predicted by \simba is consistent with the upper limits placed by the only existing
untargeted flux-limited \cii survey at the EoR and those predicted by semi-analytic models.
We compare our results with existing models and discuss the differences responsible for the discrepant slopes in
the \Lcii--\,SFR relation.
\end{abstract}
\keywords{methods: data analysis --
          galaxies: high-redshift --
          galaxies: ISM --
          galaxies: evolution --
          galaxies: formation --
          galaxies: starburst --
          stars: formation}


\section{Introduction} \label{sec:introduction}

 Deep field surveys carried out with the
 {\em Hubble Space Telescope} have enabled the detection of galaxies out to the cosmic
dawn at \z$\gtrsim\,$11 and provide stringent constraints on the bright end of rest-frame UV luminosity functions (LF)
of galaxies during the Epoch of Reionization \citep[e.g.,][]{Oesch16a, Song16a,Finkelstein16a}. 
While measuring distributions of galaxy properties, such as the
LF, provides important constraints on how galaxies evolve over cosmic time,
it is also useful to target individual sources \athighz in order to conduct detailed case studies.
Such studies allow us to address open questions such as properties of the interstellar medium (ISM), how metal enrichment in galaxies proceded in the early universe,
how much star formation is dust-obscured or how much dust is present in these early systems,
and the physical conditions under which star formation and stellar mass assembly took place.
Follow-up observations with the Atacama Large (Sub-)Millimeter Array (ALMA) probing the dust continuum and line emission from ions and molecules in the dense ISM of high-redshift galaxies appear extremely promising for characterizing the star-forming gas and dust properties of early galaxies and helping to constrain the physical processes at play (see a review by \citealt{Hodge20a}).

Owing to the relative brightness of the fine-structure line from singly ionized
carbon at rest-frame 157.7\,$\mu$m and accessibility of the line with
ALMA at high redshift, there is currently great interest in using \cii
line emission as a tracer to study galaxies during the Epoch of Reionization (EoR). At the
moment, most of these follow-up \obs target some of the brightest
objects samples selected at other wavelengths
\citep[e.g.,][]{Capak15a, Smit18a, Marrone18a}. \cii has been
detected in a handful of normal star-forming galaxies at the EoR
\citep[e.g.,][]{Capak15a, Carniani18b} and even resolved on kpc-scales
to study the gas kinematics for a handful of sources
\citep[e.g.,][]{Jones17a, Matthee17a, Carniani18b,
  Hashimoto19a}. However, due to the small field of view of ALMA, it
is extremely challenging and expensive to carry out
\emph{flux-limited} (untargeted) surveys over significant areas. The largest area untargeted survey to date that probes \cii at $z\sim 6-8$ is the ALMA Spectroscopic Survey in the HUDF (Large Program; ASPECS), which covers 4.6 arcmin$^2$ \citep{Walter16a,Aravena16a}.

While these detections are exciting, interpretation of the \cii line
luminosity and its connection with galaxy properties remains complex
and poorly understood. \cii is the dominant coolant in the ISM in
nearby star forming galaxies, and its luminosity is expected to be
correlated with the star formation rate (SFR). Indeed, such a
correlation is seen in local galaxies
\citep{DeLooze14a,Herrera-Camus15a}, but with hints that the \Lcii/SFR
ratio depends on other galaxy properties such as metallicity and dust
temperature \citep[e.g.][]{Malhotra01a, Luhman03a, Diaz-Santos14a}.
This is expected, as the dust
content affects the degree of shielding against hard ionizing photons,
as well as the amount of photoelectric heating, which affect
the ionization state of the line emitting gas and the collision rate.  In addition,
theoretical modeling has shown that \Lcii/SFR is also expected to
depend on the pressure of the ISM environment, if molecular cloud
sizes depend on the ambient pressure, and on the density distribution
on small scales within the ISM \citep{Popping19a}.

Another motivation for studying the physics of line emission and its
connection to galaxy properties and the physics of galaxy formation is
the development of multiple experiments that will carry out line
intensity mapping (LIM) studies.  Different LIM experiments will
enable the detection of various tracers of dense and more diffuse gas
in and around galaxies, including Lyman-$\alpha$, H$\alpha$, 21-cm,
CO, and [CII] (see \citealt{Kovetz17a} for a review). LIM promises to
map the statistical signal from emission in galaxies over very large
volumes (on the order of Mpc-to-Gpc-scales), with the tradeoff of not
resolving individual galaxies. As such, LIM experiments also probe
emission from faint galaxies.  While LIM holds great promise for
studying galaxy evolution and cosmology (e.g., the cosmic star
formation history, evolution of the ISM and intergalactic medium
(IGM), and physical processes during the EoR; \citealt{Kovetz17a}),
the power spectra depend on the line luminosities of different galaxy
populations within the volume sampled (or the intrinsic line LF).
Therefore, physically motivated models that can self-consistently
predict line luminosities are vital for strategizing LIM experiments
and interpreting LIM data.  As shown by \citet{Yue19a}, a shallower
\Lcii\,--\,SFR relation would imply that the \cii LF drops quickly at
the bright end and most of the IM signal comes from faint galaxies.
This in turn determines the detection depth needed for \cii LIM
experiments. At the moment, most LIM forecasts for tracers such as CO
and \cii have been made using a series of empirical scaling relations
(e.g., \citealt{Gong12a, Uzgil14a, Keating15a, Chung19a}).

Carrying out detailed and realistic predictions of the \cii emission
for a large cosmologically representative sample of galaxies is
extremely challenging.  The \cii line can arise from all phases of the
ISM, by being collisionally excited by either electrons, atoms or
molecules.  Hence, the line strength depends strongly on the density
and kinematic temperature of these species.  Modeling has repeatedly
shown that different ISM phases are all important to consider when
deriving \cii emission for a galaxy (e.g., \citealt{Olsen17a,
  Pallottini19a, Lupi20a}). However, state-of-the-art cosmological
simulations of volumes larger than 100\,Mpc do not resolve particle
masses below $\sim$\,10$^{6-7}$\,\Msun (at $z$\eq0; e.g.,
\citealt{Dave19a, Nelson18a}), which corresponds to hydrogen densities
below $n_H$\,<\,100\,\cc and temperatures above 10$^4$\,K. In
particular the molecular ISM phase --- with typical densities above a
few hundred cm$^3$ and and temperatures below 100\,K --- is not
tracked in cosmological simulations but knowledge about it is critical
in order to reliably simulate \cii line emission. Currently, all large
volume cosmological simulations adopt phenomenological ``sub-grid''
recipes to treat unresolved processes such as star formation, stellar
feedback, and chemical enrichment, and these carry significant
uncertainties \citep[see review by][]{Somerville15a}. Additionally, a
second type of ``sub-grid'' modeling is required to describe the
detailed structures of molecular gas on cloud scales, which must be
input into the radiative transfer (RT) and ionization state modeling tools
described in more detail below.

Previous works that have attempted to model \cii for galaxy
populations mainly fall into three categories: {\em (i)} extremely simple,
empirical mappings between \Lcii\ and halo mass
\citep{Visbal10a,Gong12a}, {\em (ii)} semi-analytic models to predict galaxy
properties coupled with machinery to predict the \cii emission in
post-processing \citep{Popping16a,Lagache18a,Popping19a}, or {\em (iii)} small
samples of cosmological zoom-in simulations again post-processed to
compute the radiative transfer and line emission \citep{Olsen15a,
  Narayanan15a, Vallini15a, Olsen17a, Katz17a, Pallottini17a,
  Pallottini17b}. A few recent studies have coupled on-the-fly
radiative transfer, non-equilibrium chemistry modeling, and line spectral
synthesis with ultra high resolution hydrodynamic zoom-in simulations
\citep{Katz17a,Katz19a,Pallottini19a}.

A variety of tools are brought to bear to compute the emergent line
emission in the literature. A stellar population synthesis code such
as \ncode{Starburst99} is used to derive the amount of intrinsic
radiation from stellar emission \citep{Leitherer14a}. Codes such as
\ncode{Skirt} or \ncode{Powderday} are used to perform dust RT
\citep{Camps15a, Narayanan15a}, and tools such as \ncode{RADMC-3D},
\ncode{Despotic}, \ncode{Lime}, \ncode{Cloudy}, or \ncode{MAPPINGS}
are used to compute line RT in the ISM \citep{Allen08a, Krumholz14a,
  Ferland17a}.  For a detailed summary and comparison of these codes,
we refer interested readers to \citet{Olsen18a}.

Some of the biggest differences among the aforementioned codes are the
density range considered/permitted, the geometry, and the species
included in the chemical network.  These differences also imply
different demands on computational time and memory, each with
different benefits and trade-offs.  The accuracy needed for a given
galaxy simulation and the emission line of interest typically determine the
method used. For instance, a photo-dissociation region (PDR) code such
as \ncode{Despotic} can be used to calculate line emission from the
neutral ISM; however, \ncode{Despotic} does not simulate line emission
originating in the ionized phase of the ISM.

In addition to different methods being adopted for RT and line
spectral synthesis, the type of simulation used also sets limits on
the galaxy sample size obtainable and the level of realistic physics
that can be adopted.  SAMs are computationally inexpensive and can
easily generate galaxy catalogues of statistically significant sample
sizes (e.g., \citealt{Popping16a, Lagache18a} and
\citealt{Popping19a}), and thus are excellent for testing physical
recipes and exploring wide ranges of parameter space, but they do not
provide any detailed information on sub-galactic structure. In
contrast, the highest resolution zoom-in hydrodynamical simulations can
numerically resolve the ISM down to scales of $\sim$ 10\,pc at high
redshifts (e.g., \citealt{Katz17a, Pallottini17b, Pallottini19a}), but
are computationally expensive. The samples of
galaxies with simulated line emission based on numerical
hydrodynamical simulations are thus limited in number to 1-30 per
study, and therefore also probe a limited parameter space of galaxy
properties. Previous studies commonly targeted the most massive halos
and/or a handful of sources with properties resembling some known
properties of a given observed $z$\ssim6 galaxy.  Over the years, the
resolution of large volume cosmological hydrodynamics simulations has
increased significantly, reaching $\gtrsim$100\,pc resolution even
before carrying out additional refinement using zoom-in
techniques. This produces a statistically significant and unbiased
sample of galaxies while reaching down to a relevant spatial scale to
simulate emission emerging from the ISM (though sub-grid models are
still required). \cite{Inoue20} gave a successful demonstration of calculating CO line emission in direct post-processing of the cosmological IllustrisTNG simulation using a simulation box size of 75\,Mpc.

In this work, we leverage the new \simba suite of cosmological
hydrodynamic simulations (\citealt{Dave19a}) to select a galaxy sample
at $z\sim 6$ for line emission postprocessing that is unprecedented in
its size (11,137 galaxies) and dynamic range (halo mass $\sim
10^9-10^{12} \Msun$; stellar mass $\sim 10^7-10^{11} \Msun$). The
sample is drawn from a set of volumes that vary in resolution, such
that the sample from each volume is representative of the population
that is well resolved in the simulation. We apply and updated version
of the \sigame package \citep{Olsen17a}, which includes
sub-resolution modeling of the ISM, radiative transfer and line
spectral synthesis. We use this calculation to provide predictions of
how the \cii luminosity at this redshift is related to other
galaxy and DM halo properties, and of the \cii LF, and compare our results with
available observations and with other models in the literature.

The paper is structured as follows.  In \Sec{sim}, we describe the
\simba simulations, and in \Sec{sigame}, the method used to
simulate \cii line emission.  We present the results in \Sec{results}
and discuss the limitations in \Sec{caveats}.  Finally, conclusions
and implications of our findings are presented in \Sec{conclusion}.
Throughout this paper, we adopt a concordance cosmology, with total
matter, vacuum and baryonic densities in units of the critical density
$\Omega_{\Lambda}$\eq0.693, $\Omega_m$\eq0.307, $\Omega_b$\eq0.048, a
dimensionless Hubble parameter $h$\eq0.678, scalar spectral index of
$n$\eq0.96 and power spectrum normalization of $\sigma_8$\eq0.823
\citep{Planck16a}.

\section{Simulation}\label{sec:sim}

\subsection{Cosmological Hydrodynamics Simulation: \ncode{simba}} \label{sec:simba}
We use galaxies from the \simba cosmological galaxy formation simulations \citep{Dave19a} for this study.
\ncode{simba} is a suite of \ncode{gizmo}-based simulations using meshless finite mass
hydrodynamics,
incorporating state-of-the-art feedback modules that provide very good agreement
with a wide range of lower-redshift observables.
The suite consists of random cubical volumes of 100, 50, and 25\,\cmpc\ on a side, each with
1024$^3$ dark matter particles and 1024$^3$ gas elements.  By combining the results
of these simulations, we achieve unprecedented dynamic range, with the highest resolution
equalling e.g. that in a study of far-infrared lines using zoom simulations presented
in \citet{Olsen17a}.
The smoothing lengths and the initial gas mass resolutions for the 100, 50, and 25\,\cmpc\ volumes
are $\epsilon_{\rm min}$\eq0.5, 0.25, and 0.125\,$h$\pmOne\,kpc
and $m_{\rm gas}$\eq1.8\E{7}, 2.3\E{6}, and 2.9\E{5}\,\Msun, respectively
(see Table 1 of \citealt{Dave19a}).
These runs all use identical input physics, begin at
$z$\eq249 and assume a Planck-concordant cosmology.

\simba is the successor of the \ncode{mufasa} simulations (\citealt{Dave16a}), and details
of the improvements in \simba are provided in \citet{Dave19a}.
Among the various updates, key ones relevant to this work are:
{\em (i)} \simba uses the \ncode{grackle-3.1} package
to model radiative cooling and photoionization heating, updated from \ncode{mufasa}
to apply radiative processes via isochoric substep cycling, and also computing
the neutral hydrogen content accounting for self-shielding
on-the-fly via the prescription in \citet{Rahmati13a};
{\em (ii)} \simba explicitly models the growth and feedback of
supermassive black holes (SMBH) residing in galaxies, with
the growth of the SMBH set by
torque-limited accretion of cold gas~\citep{Angles17a} and
Bondi accretion of hot gas, while black hole feedback
is modeled via bipolar kinetic outflows 
and injection of X-ray energy;
{\em (iii)} \simba includes a sub-grid model to form and destroy dust within the ISM of galaxies during the simulation run,
with the dust mainly produced by Type II supernovae, asymptotic giant branch stars
and condensation from metals, and destroyed
predominantly via sputtering (including supernova shocks) and consumption by \SF
\citep{Li19a};
{\em (iv)} \simba employs ejective star formation feedback like \ncode{mufasa}, but with scalings updated to reflect particle tracking results from the
Feedback in Realistic Environment (FIRE) simulations~\citep{Angles17b},
with minor modifications to better reproduce EoR galaxy properties
(See \citealt{Wu19a}, under review for details).

\simba has been compared to a wide range of observations across cosmic time, and is found to
provide reasonable agreement, including for the galaxy stellar mass function (GSMF)
and mass metallicity relation \citep{Dave19a}, black hole properties \citep{Thomas19a},
dust properties \citep{Li19a}, galaxy sizes and profiles \citep{Appleby2020},
and cold gas contents including \aco luminosity functions
from $z=0-2$ \citep{Dave20a}.  Minor disagreements with \obs include an overproduction
of the very highest mass galaxies at $z\la 2$, too-large size for low-mass quenched
galaxies at \z$\la 1$, and an underproduction of the dust mass function at \z$\sim 2$.
Relevant to this work where we focus on $z=6$,
Wu et al., (2019, under review) examined the EoR properties of \simba galaxies
and found good agreement at $z=6$ with the UV luminosity function, UV slope measurements when a
\citet{Calzetti01a} dust model is assumed, and galaxy sizes down to the faintest limits.
Hence \simba provides a realistic platform to examine the far-infrared line properties
of EoR galaxies, which is the goal of this study.

\begin{figure}[htbp]
\centering
\includegraphics[trim=0 0 0 28, clip, width=.5\textwidth]{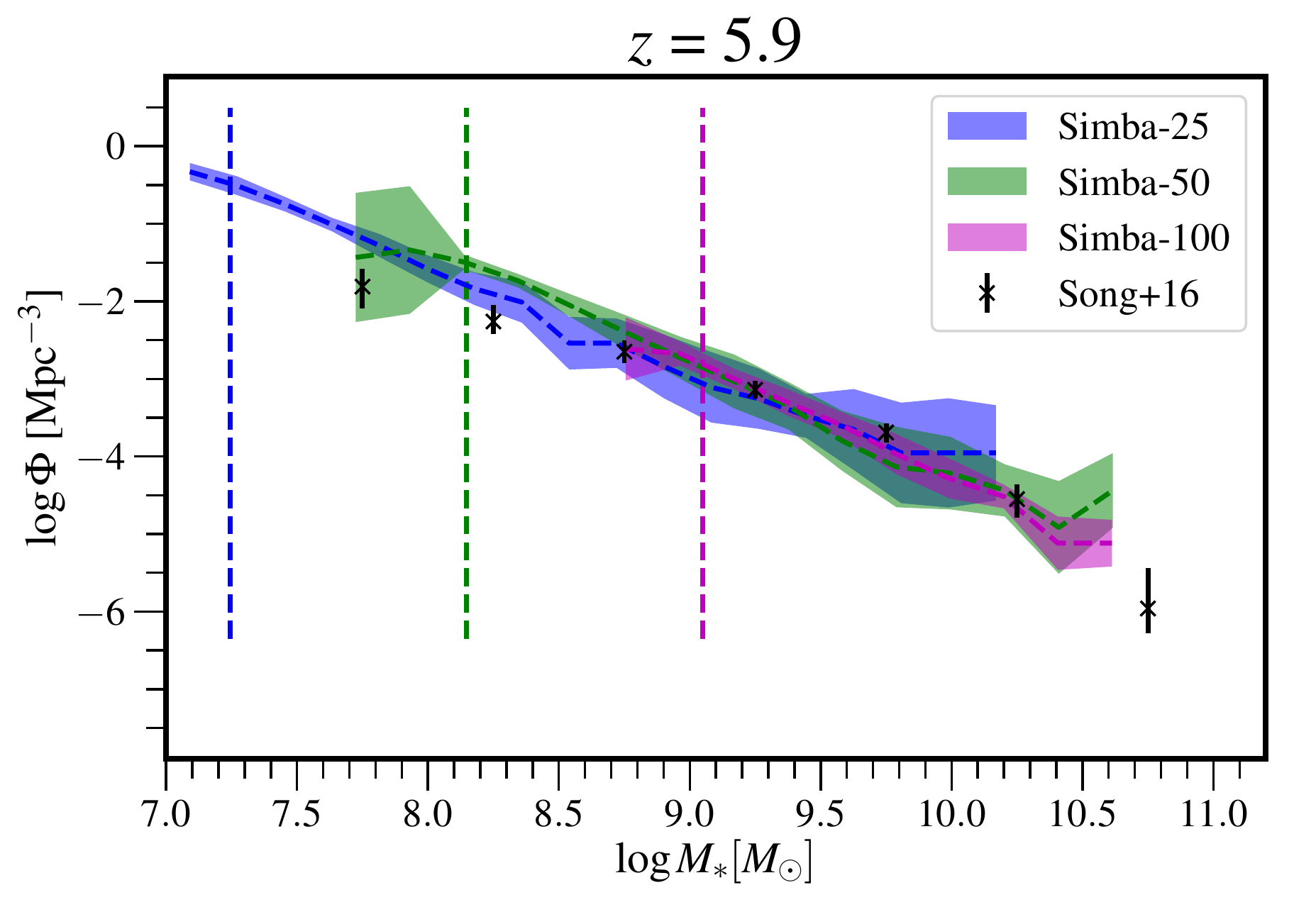} \\
\includegraphics[trim=0 0 0 0, clip, width=.5\textwidth]{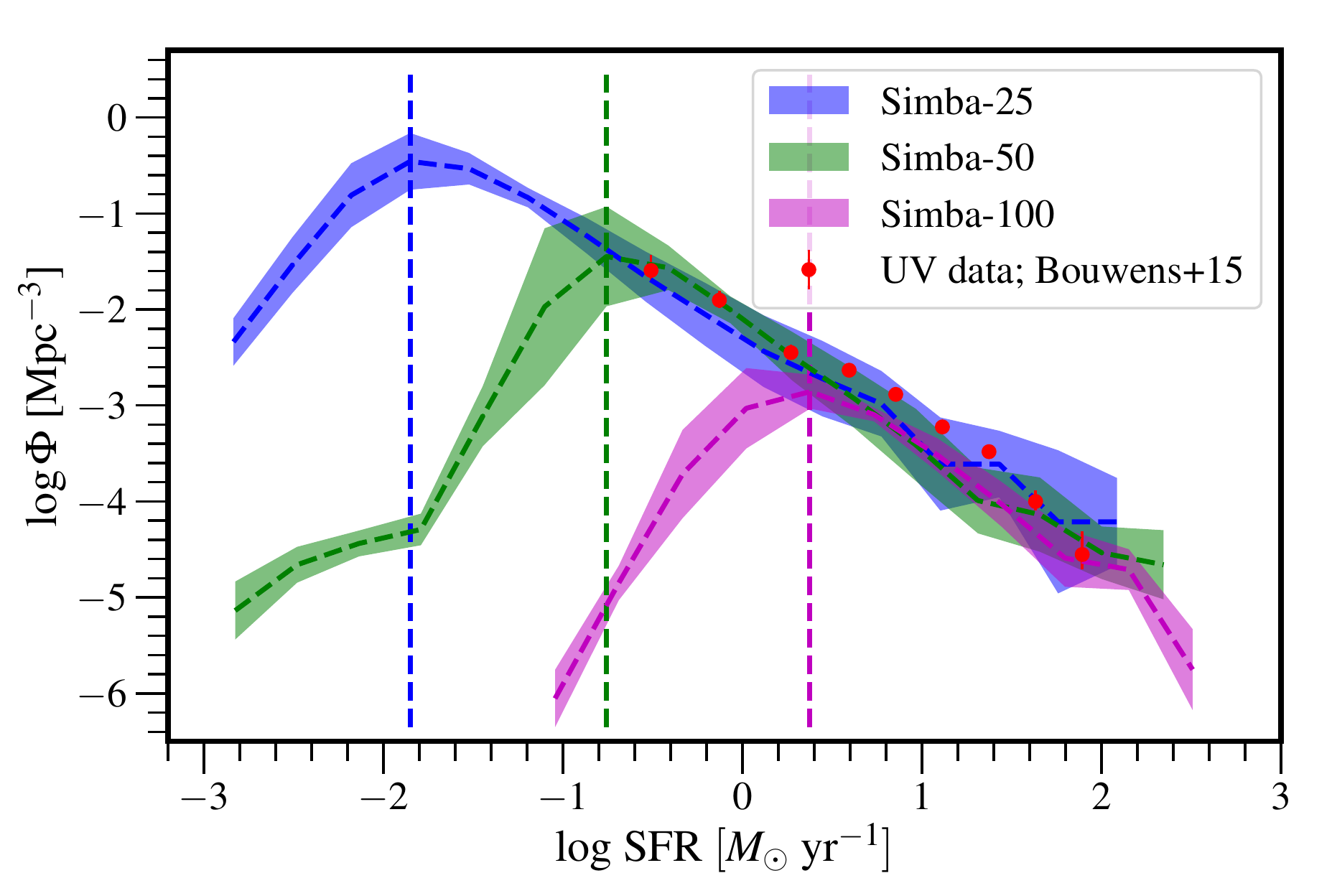}
\caption{
Top: Galaxy stellar mass function for the different simulation boxes at $z$\eq6 compared to
the results based on a rest-frame UV selected observational sample (black markers).
Vertical dashed lines show the mass requirement applied to each of the simulation boxes (color-coded)
to select only galaxies that are resolved by the simulation (see \Sec{simba}).
Bottom: Same as the top panel, but for the SFR function.
Vertical dashed lines show the thresholds in SFR for each box, below which the boxes become incomplete.
The turnover arises from the finite mass resolution of these simulations
(before applying the selection criteria; see \Sec{sample}).
Observational results (corrected for dust attenuation) are plotted as red symbols.
The spread marked by the shaded regions is computed from jackknife resampling eight sub-octants of the simulated volumes.
Observations are from \citet{Song16a} (top) and \citet{Bouwens15a} (bottom). The \simba predictions are in very good agreement with the observational estimates.
\label{fig:GSMF}}
\end{figure}

In \Fig{GSMF}, we show the GSMF and SFR function at $z$\,\ssim6.
We point out the robust numerical convergence in the GSMF and SFR function
(and galaxy properties, see \citealt{Dave19a} and \Sec{sample}) without
refining or fine-tuning the parameters in the sub-grid models of \ncode{simba}
between the various simulation boxes at the redshift studied in this work.
Note that we do not impose any stellar mass limits in the GSMF plotted.
This illustrates the exceptional convergence reached across the boxes.
This is crucial as we make predictions of the \cii LF in the luminosity range of 5.5\,$<\log$\,(\Lcii/\Lsun)\,$<$\,8.5 by combining galaxies in the $25 \cmpc$, $50 \cmpc$, and $100 \cmpc$
boxes (hereafter Simba-25, Simba-50, and Simba-100, respectively).

\subsection{Main Sample: 11,137 Galaxies at \z$\simeq$\,6} \label{sec:sample}
Galaxies from the simulation suite are identified using a galaxy finder that adopts a
6-dimensional friends-of-friends algorithm (\ncode{caesar}).
For the purpose of this work, we include only galaxies that have at least 64 stellar and gas particles,
respectively, to ensure they are resolved in the simulation.
As illustrated in \Fig{GSMF}, these mass requirements correspond to
$\log {(M_{\rm \star,min}/M_\odot)}$\eq7.24 for Simba-25,
$\log {(M_{\rm \star,min}/M_\odot)}$\eq8.15 for Simba-50, and
$\log {(M_{\rm \star,min}/M_\odot)}$\eq9.05 for Simba-100.
Similarly, we impose thresholds on the SFR averaged over 10\,Myr based on the
turnover seen in the SFR function indicating incompleteness in SFR.
This corresponds to $\log\left(\textrm{SFR}/\textrm{M}_\odot\,\textrm{yr}^{-1}\right)>-$1.9, $-$0.8, and $0.4$
for Simba-25, -50, and -100 respectively.
In addition, we only include galaxies with a molecular gas mass of at least
$M_{\rm mol}$\eq$f_{\rm H2}\,M_{\rm gas}$\,$>$10$^5$\,\Msun, as the
sub-grid model has to form giant molecular clouds (GMCs) of at least 10$^4$\,\Msun each by
sampling the cloud mass from a GMC mass function (see \Sec{sigame}).
After applying these criteria, we have a sample of $N_{\rm tot}$\eq11,137 galaxies for a single snapshot at $z$\,\eq6.
The ranges of their physical properties are listed in \Tab{param}.

\begin{figure}[htbp]
\centering
\includegraphics[trim=0 0 0 0, clip, width=.5\textwidth]{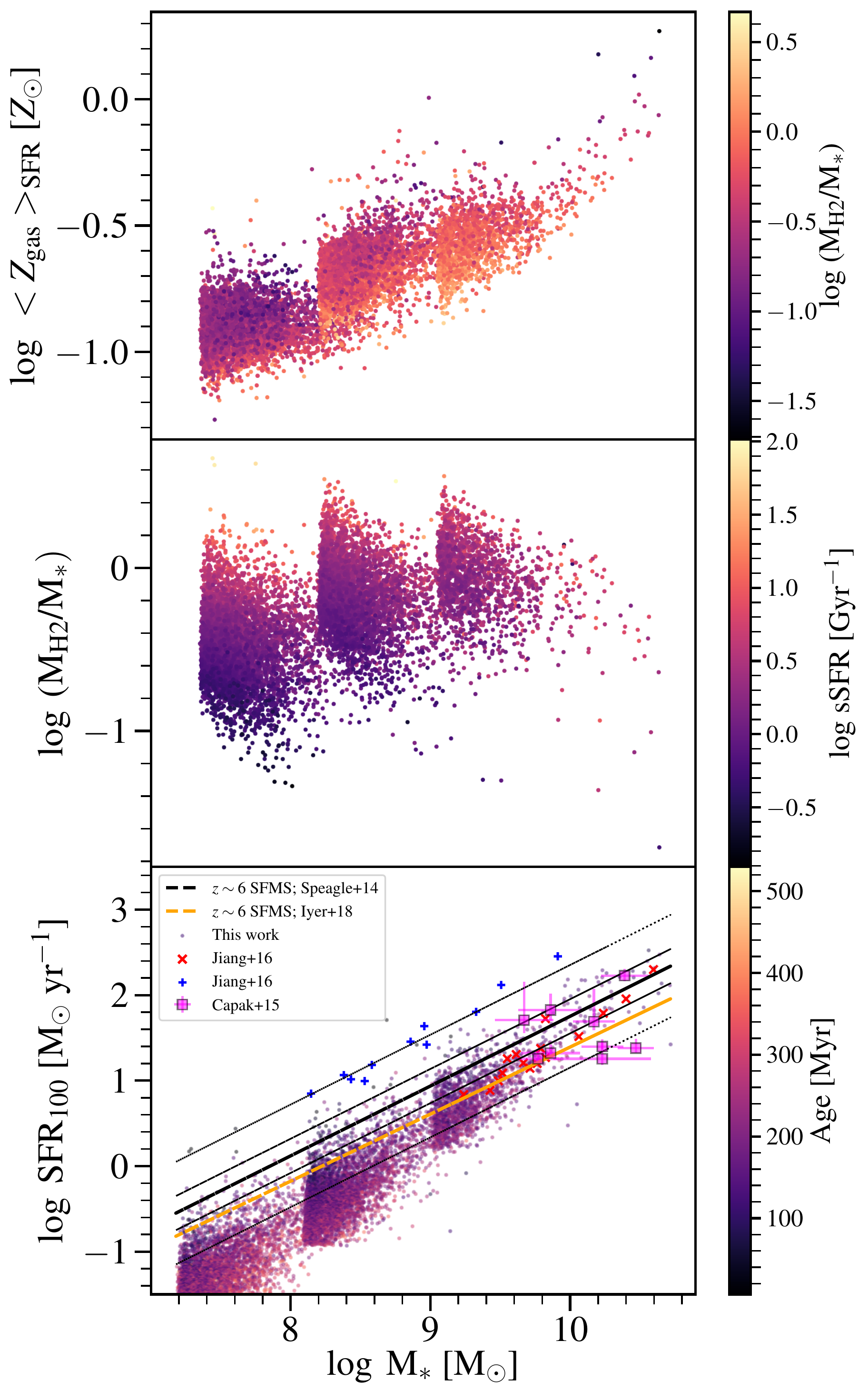}
\caption{
Scaling relations for the \simba galaxy sample (circular dots).
Top: $\langle Z_{\rm gas}\rangle_{\rm SFR}$ -- \mstar relation, color-coded by the molecular gas mass fraction.
Middle: Molecular gas-to-stellar mass ratio ($M_{\rm H2}$/\mstar) --  \mstar relation, color-coded by sSFR.
Bottom: SFR--\mstar relation, color-coded by the mass-weighted stellar age.
 The SFR of the simulated galaxies are averaged over 100\,Myr.
Magenta squares, red crosses, and blue plus symbols
correspond to \obs of UV-selected star-forming galaxies at \z$\sim$\,6 \citep{Capak15a, Jiang16a},
with the red markers indicating older galaxies with a crude estimated age of
$\gtrsim$\,100\,Myr and 
the blue ones indicating younger galaxies with age $\lesssim$\,30\,Myr.
Dashed lines correspond to empirical relations for
the star-forming main sequence (SFMS) and their 1\sig\ and 3\sig\
spreads at this redshift \citep{Speagle14a, Iyer18a}.
The sharp cutoffs seen in the last two plots results from the mass cut imposed on each of the simulation boxes
(Simba-25, Simba-50, and Simba-100) to only include galaxies that are numerically resolved (see \Sec{sample}).
\label{fig:prop}
}
\end{figure}

To illustrate the range of physical properties sampled by the \simba galaxies,
\Fig{prop} shows the relations between the specific SFR (sSFR),
SFR-weighted metallicity ($\langle Z_{\rm gas}\rangle_{\rm SFR}$), molecular gas-to-stellar mass ratio ($M_{\rm mol}$/\mstar),
and the stellar mass-weighted age in our $z$\eq6 sample.
The three clumps of points represent galaxies from our three simulation volumes, and generally
show reasonable convergence.
The weighted quantities are indicated with the $\langle...\rangle$ notation (e.g., $\langle Z_{\rm gas}\rangle_{\rm SFR}$), as
defined as follows:
\begin{equation}\label{eqn:defineaverage}
\langle x \rangle \equiv \frac{\sum_{i} \rho_i x_i }{\sum_i \rho_i}\,,
\end{equation}
where $x$ is the variable and $\rho_i$ is the volume of each fluid element $i$.
The $\Sigma_{\rm SFR}$ in this paper is defined as:
\begin{equation}
\Sigma_{\rm SFR} = \frac{\textrm{SFR}}{\pi R_{\rm 1/2}^2}
\end{equation}
where $R_{\rm 1/2}$ is the half-mass radius of gas of a given galaxy.
The scaling relations shown are similar to the mass-metallicity
and fundamental metallicity relation (a.k.a. MZR and FMR; e.g., \citealt{Maiolino08a, Mannucci10a}),
which are commonly used to gain insights into the interplay between \SF, gas accretion, and
feedback during the evolution of a galaxy,
and are shown to illustrate the range of physical properties sampled by the \simba galaxies.
As can be seen in the second panel, most of the \simba galaxies are rich in molecular gas.
The trend of decreasing $M_{\rm mol}/$\mstar with increasing \mstar
arises owing to stellar feedback which preferentially suppresses the stellar mass in lower
mass systems, thereby increasing
$M_{\rm mol}$/\mstar.
The middle panel also shows how, for a given stellar mass bin,
the SFR increases with the molecular gas mass fraction, as expected.
That said, there are certainly jumps in $M_{\rm mol}/$\mstar fractions between the different simulations volumes indicating less than ideal convergence, although this becomes less apparent when plotting $M_{\rm mol}$ against \mstar as shown in the left panel of Fig.\,\ref{fig:mH2} in the Appendix.
The bottom panel displays the \simba galaxies on top of the  ``star-forming main sequence'' (SFMS; e.g., \citealt{Speagle14a, Iyer18a}),
which most of our galaxies follow at $\log M_*\gtrsim9$\,\Msun --- which is also
the stellar mass limit of the observational data. At lower stellar mass, the sSFR
of \simba galaxies falls below the SFMS extrapolation\footnote{Such extrapolation
assumes that the SFMS follows the same power law as that at the high mass end, which is not directly observed.}.
The color coding in the three panels shows that, at a given stellar mass,
galaxies that are higher metallicity have lower gas contents, galaxies
that have higher gas contents have higher sSFR, and galaxies that have higher
sSFR have lower mean stellar age.

Galaxies of similar stellar masses and SFRs may have different sizes,
surface densities, gas contents, metallicities, interstellar radiation field strengths,
structural properties, and gas dynamics;
all of which would produce varying \cii luminosities (see e.g., \citealt{Kaufman99a, Vallini15a, Olsen17a}).
As such, comparing the physical properties of observed (i.e., \cii-detected ones in the context of this work) and
simulated galaxies is pertinent to establishing the reliability of model predictions.
In other words, comparing these global properties of \ncode{simba} galaxies with those of observed galaxies
enables one to place the observed ones, given their \cii luminosities,
in a theoretical framework.
It is beyond the scope of this paper to perform a detailed comparative study since the information
available for observed galaxies at these redshifts (see \Sec{ciifir}) remains limited and inhomogeneous.
For comparisons of \simba galaxies with \obs at lower redshifts and discussion on
redshift evolution in these relations, we refer interested readers to \citet{Dave19a}.
At the EoR, the size-luminosity relation of \simba galaxies agrees with \obs
(\citealt{Kawamata18a, Wu19a}).

\input{Table_paramSpace.tex}

\section{Method: Simulating Line Emission} \label{sec:sigame}

We use an updated version of \sigame \citep{Olsen15a, Olsen17a}
to post-process the \simba simulation outputs.
For details of the code, we refer interested readers to \citet{Olsen17a}.
Here, we briefly summarize the salient points of \sigame and updates made to the code as part of this work.
For each gas fluid element, \sigame divides the molecular gas mass (i.e., $f_{\rm H2, i}\,m_{\rm gas, i}$) into
GMC by sampling the Galactic GMC mass function
over the mass range of 10$^{4-6}$\,\Msun
($dn/dM\propto M_{\rm GMC}^{-1.8}$; see e.g., the review by \citealt{McKee07a, Blitz07a, Kennicutt12a}).
The remaining mass of the parent fluid element is assumed to be in the diffuse gas phase, and
is subsequently distributed into diffuse ionized and neutral gas phases. This division is determined
by the boundary at which the inner neutral region transitions to the outer ionized region of the diffuse clouds as computed from RT calculations within \ncode{Cloudy}.
That is, the neutral gas phase corresponds to the region beyond a radius where the neutral fraction $x_{\rm HI} = n_{\rm HI} / (n_{\rm HI} + n_{\rm HII}) > \,$0.5, such that it is dominated by neutral hydrogen --- here $n$ is the number density.
As such, \sigame accounts for
line emission from three distinct ISM phases.
The smoothing length of the parent fluid element is adopted as the size of the diffuse gas clouds, whereas the size of each GMC
 is derived from a pressure-normalized mass-size relation, following
\begin{equation}
\frac{R_{\rm GMC}}{\rm pc} = \left(\frac{P_{\rm ext}/k_{\rm B}}{10^4\,{\rm cm}^{-3}\,{\rm K}}\right)^{-1/4}\left(\frac{M_{\rm GMC}}{290~M_\odot}\right)^{1/2},
\end{equation}
where $k_{\rm B}$ is the Boltzmann constant
and the external cloud pressure $P_{\rm ext}$ is defined assuming mid-plane hydrostatic equilibrium within the galaxy.

Each GMC radial density profile is assumed to follow a truncated logotropic profile.
Both GMCs and diffuse gas phases inherit the metallicity of their parent fluid element.
The FUV luminosity of each star is calculated based on its age and metallicity
and is determined by interpolating over a grid of \ncode{starburst99}
stellar population synthesis models \citep{Leitherer14a} (with the default \citealt{Kroupa02a} initial mass function).
Each GMC is irradiated by a local FUV radiation (6--13.6\,eV), where the strength
of the radiation field ($G_0$ in Habing units; 1.6\E{-3}\,erg\,cm$^{-2}$\,s\,\pmOne)
is determined by summing up the FUV flux from nearby stellar particles and by assuming that the flux falls
off as $1/r^2$.
In the diffuse gas phase, the FUV radiation field is determined based on the SFR surface density
of the galaxy.
For our sample, the SFR surface density ranges between $\simeq$\,1\,--\,6200 times the Milky Way.

We use the photoionzation code \ncode{cloudy} version 17.01 \citep{Ferland17a}
to simulate the thermo-chemistry in the three distinct ISM phases tracked by \sigame
by performing detailed balance calculations of the various species, taking into account physical
processes such as H$_2$ photo processes, dust physics (grain-atom/ion charge
transfer), 
and cosmic ray (CR) ionization.
The line luminosities are then derived from the cooling rates for different line transitions.
For computational purposes, lookup tables are generated for the
GMC and the diffuse (neutral and ionized) gas phase, respectively.
The FUV radiation field impinging on the gas phases
is assumed to have the same spectral shape as in the solar neighborhood.

Cosmic rays are added, with an ionization rate equal to that of the Milky Way scaled linearly by a
factor of $(G_{0, \rm gas}/G_{0, \rm MW})$.
For the GMC models, the clouds are in theory completely embedded within diffuse gas,
and thus H-ionizing radiation is turned off in the \ncode{Cloudy} models (cf. \citealt{Olsen17a}).

The main parameters in the GMC phase of the
\ncode{cloudy} models considered are the $G_{\rm 0, GMC}$ of the radiation source,
radius of the cloud ($R_{\rm GMC}$), and cloud density profile as a function of cloud radius ($n_H(R_{\rm GMC})$).
Turbulent velocity is added to the GMC models according to the velocity dispersion calculated from the
cloud radius and pressure, assuming clouds are virialized.
For the diffuse gas phase, the main model parameters are
gas density ($n_H$), gas kinetic temperature ($T_k$), diffuse cloud size,
($R_{\rm dif}$), and metallicity ($Z$).

Compared to \citet{Olsen17a}, the main updates made to \sigame used in this work are as follows:
\begin{itemize}
\item Instead of fixing the number and width of shells used by \ncode{cloudy} to model each GMC,
we now allow \ncode{cloudy} to determine the optimal quantities to ensure convergence.
This modification leads to more accurate calculation of the
grain photoelectric heating of the gas and increases the importance of gas heating
due to this mechanism in the GMC models, which
is the main excitation mechanism for \cii emission.
Namely, \cii is collisionally excited such that higher kinetic temperature leads to more molecular motions and collisions
inside GMCs and photo-dissociation regions (PDRs), the main sites for \cii emission in galaxies.
\item To ensure good sampling of the parameter space for both GMCs and diffuse gas clouds, the
number of \ncode{cloudy} models used to create look-up tables is significantly increased
from 1296 to 4096 models by using 8 grid points in each parameter space dimension rather than 6
as in \citet{Olsen17a}. The look-up tables are further described in \Sec{grids}.
\item The dust content of the ISM is a crucial factor in setting the \cii luminosity.
As often done, we will assume here that dust scales with metallicity via the dust to metal ratio (DTM), but
instead of using a solar DTM value of $\sim0.46$ (as done in \cite{Olsen17a}),
we take a DTM of 0.25 based on the mean value of our \simba galaxies (see \Sec{grids}).
\end{itemize}

\subsection{GMCs and Diffuse Gas Phase Properties and \ncode{cloudy} Model Grids}  \label{sec:grids}

As mentioned in the previous section, the parameters passed to \ncode{cloudy}
for GMCs and diffuse gas clouds are;
$M_{\rm GMC}$, $G_{\rm 0, GMC}$, $Z$, and $P_{\rm ext}$ for the GMC models
and in $n_H$, $T_k$, $R_{\rm dif}$, and $Z$ for diffuse gas models.

For the GMCs, we generate 4096 models spanning 
$\log (M_{\rm GMC}/M_\odot)\in$[4.1, 4.3, 4.6, 4.8, 5.1, 5.3, 5.6, 5.8],
$\log (G_{\rm 0, GMC}/G_{\rm 0, MW})\in[$0.3, 1.0, 1.6, 2.3, 3.0, 3.7, 4.3, 5.0],
$\log (Z/Z_\odot)\in[-$3, $-$2.5, $-$2.1, $-$1.6, $-$1.2, $-$0.7, $-$0.3, 0.2], and
$\log (P_{\rm ext}/k_B)\in[$4.0, 4.9, 5.7, 6.6, 7.4, 8.3, 9.1, 10.0]\,cm$^{-3}$\,K.
For the diffuse gas, we first determine the SFR surface density of all the galaxies of the sample.
We then define a range of FUV grids over $(G_{\rm 0, GMC}/G_{\rm 0, MW})\in$[0.8, 7.2, 68, 650, 6200]
based on the ranges of SFR surface densities found in the simulated galaxies as a hyperparameter.
For each of the FUV grid, we generate 4096 models spanning
$\log (n_H/{\rm cm}^3)\in$[$-$5.0, $-$4.32, $-$3.66, $-$2.99, $-$2.31, $-$1.64, $-$0.97, $-$0.3],
$\log (T_k/{\rm K})\in$[2.5, 3.0, 3.6, 4.2, 4.8, 5.4, 5.9, 6.5],
$\log (R_{\rm dif}/{\rm kpc})\in$[$-$0.7, $-$0.49, $-$0.27, $-$0.06, 0.16, 0.37, 0.59, 0.8], and
$\log (Z/Z_\odot)\in[-$1.0, $-$0.83, $-$0.66, $-$0.49, $-$0.31, $-$0.14, 0.03, 0.2].

The gas kinetic temperature in the GMC phase is left as a free parameter to be determined by solving the thermal balance equation
in \ncode{cloudy},
whereas in the diffuse gas phase, the temperature is fixed to the grid points representative of the range seen in
the gas fluid elements from the \simba simulation.
The effect of gas heating due to the photo-excitation by the cosmic microwave background (CMB)
at the EoR is included. The resulting line intensities
are corrected to give the net flux above the background continuum
(i.e., not the contrasting flux that observers would measure; see e.g., \citealt{daCunha13a}),
and include the diminution effect where the upper levels are sustained by CMB
(\citealt{Ferland17a}).

\subsection{Dust and Elemental Abundances} \label{sec:element}
While \simba tracks dust in the simulation, we do not create a different \ncode{cloudy} lookup table for
each dust-to-mass (DTM) ratio found in the simulated galaxies since this becomes computationally intractable (i.e., it corresponds to
a hyperparameter where each DTM ratio would have a separate set of 4096 \ncode{cloudy} models).
Instead, we adopt a DTM ratio based on the median of \simba galaxies at $z\sim6$, corresponding to $\xi_{\rm DTM}$\eq0.25, which is defined as
\begin{equation}
\xi_{\rm DTM} = \frac{M_{\rm dust}}{f_Z~M_{\rm gas} + M_{\rm dust}},
\end{equation}
where $M_{\rm dust}$ and $M_{\rm gas}$ are the dust mass and total gas mass in solar mass units, and
$f_Z$ is the mass fraction of metals (i.e., $f_Z~M_{\rm gas}$ yields the mass of metals in gas-phase).
The dust content of each cloud is then set to scale linearly with its metallicity through this DTM.
The default set of lookup tables in \ncode{Cloudy} assumes a DTM of $\xi_{\rm DTM}$\eq0.46 at solar metallicity.
A more commonly adopted expression for the DTM is:
\begin{equation}
\textrm{DTM} \equiv \frac{\textrm{DGR}}{\textrm Z}.
\end{equation}
In the Milky Way, Z\eq$Z_\odot$ and DGR is $\sim$0.01, yielding $\log$~DTM\eq$-$2.

In \ncode{cloudy}, one can supply the total metallicity, and the abundance of each of the elements is scaled correspondingly
assuming Solar composition (i.e., abundance ratios of the Sun).
In order to account for abundance patterns of galaxies that differ from Solar,
we use the abundance of each metal tracked in \simba (i.e., He, C, N, O, Ne, Mg, Si, S, Ca, and Fe).
For elements that are not tracked in the simulation, we use Solar abundance ratios.
Since the mass fraction of each element tracked varies as a function of metallicity, we
fit a spline curve to the running means across all gas fluid elements.
This provides a function that maps a given metallicity to an abundance pattern, such that
the relative elemental abundances in the \ncode{cloudy} input are scaled according to the
metallicity of the cloud. In the Appendix we show the particle distribution and scalings found for our sample of \simba galaxies. We note that some elements, like carbon and nitrogen, can be quite far from their solar abundance value, which in \simba is a result of including enrichment from Type II supernovae (SNe), Type Ia SNe, and Asymptotic Giant Branch (AGB) stars, with separate yield tables for each class of star as described in \cite{Oppenheimer2006}.

\section{Results and Discussion}\label{sec:results}
\subsection{\cii -- SFR Relation at \z$\simeq$\,6} \label{sec:ciifir}

\begin{figure*}[htbp]
\centering
\includegraphics[trim=0 0 0 0, clip, width=.9\textwidth]{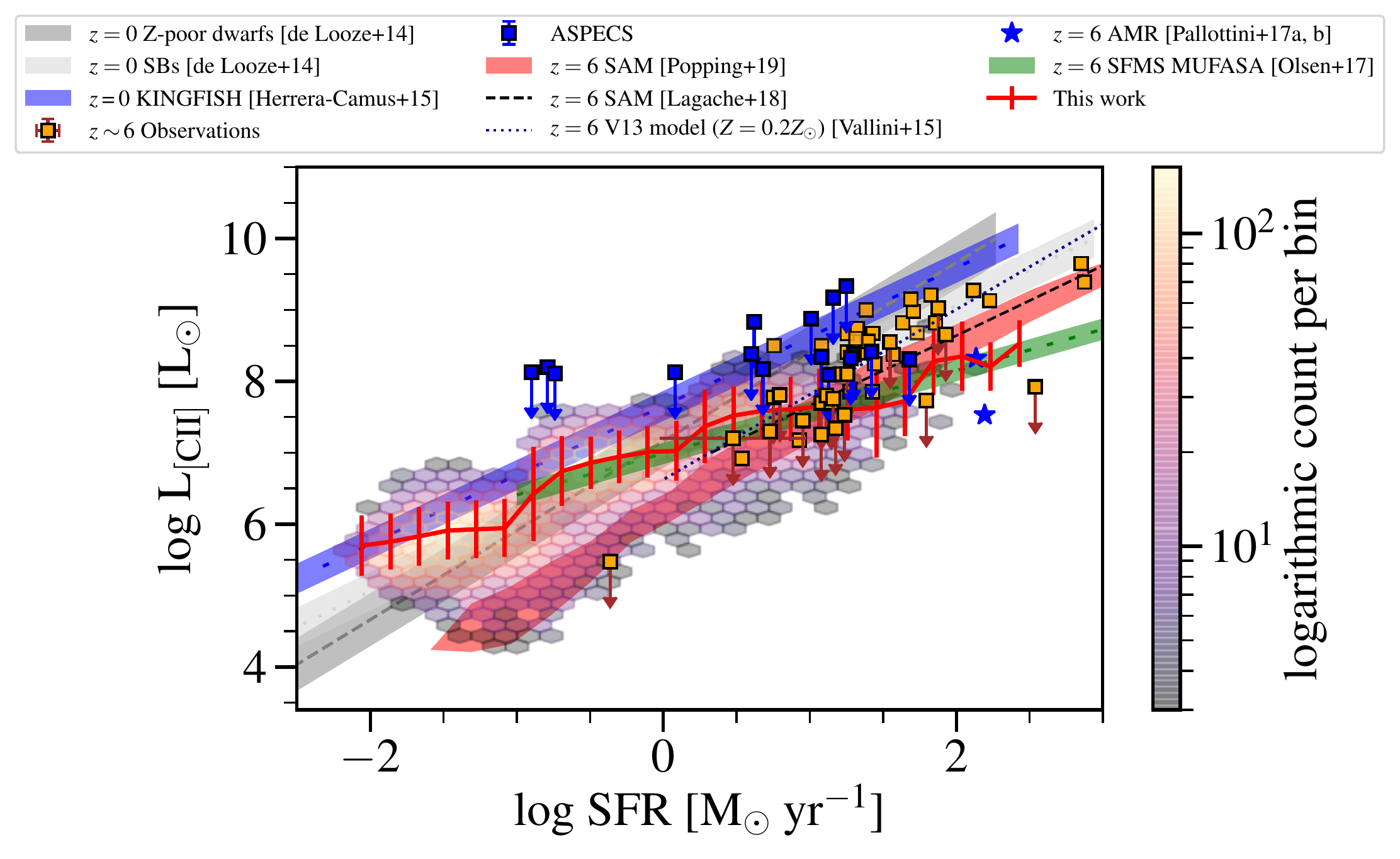}
\caption{SFR and \Lcii of \simba galaxies at \z\eq6 (hexbin) compared to existing \obs and models at the EoR.
Red lines show the running mean and standard deviations of the binned data for \simba galaxies.
Results from a sample of 30 (zoom-in)
\mufasa galaxies at $z$\,$\simeq$\,6 are shown as green shaded regions \citep{Olsen17a},
whereas those from SAM-based predictions at \z\eq6 are shown as light red shaded regions \citep{Popping19a},
and results from zoom-in AMR simulations from \citet{Pallottini17a} and \citet{Pallottini17b} are shown as blue stars.
Fits to observations from $z$\,\eq0 are shown as gray and blue
shaded regions \citep{DeLooze14a, Herrera-Camus15a}.
Square symbols show observations at $z\simeq$6 compiled from \citet{Ouchi13a, Kanekar13a,
Ota14a, Gonzalez-Lopez14a, Maiolino15a,
Schaerer15a, Capak15a, Willott15a,
Bradac17a, Inoue16a, Pentericci16a,
Knudsen16a, Knudsen17a, Decarli17a, Smit17a, Carniani18a} and Uzgil et al. 2020, in prep.
\label{fig:ciisfr}}
\end{figure*}

\begin{figure}[htbp]
\centering
\includegraphics[trim=0 20 0 0, clip, width=.55\textwidth]{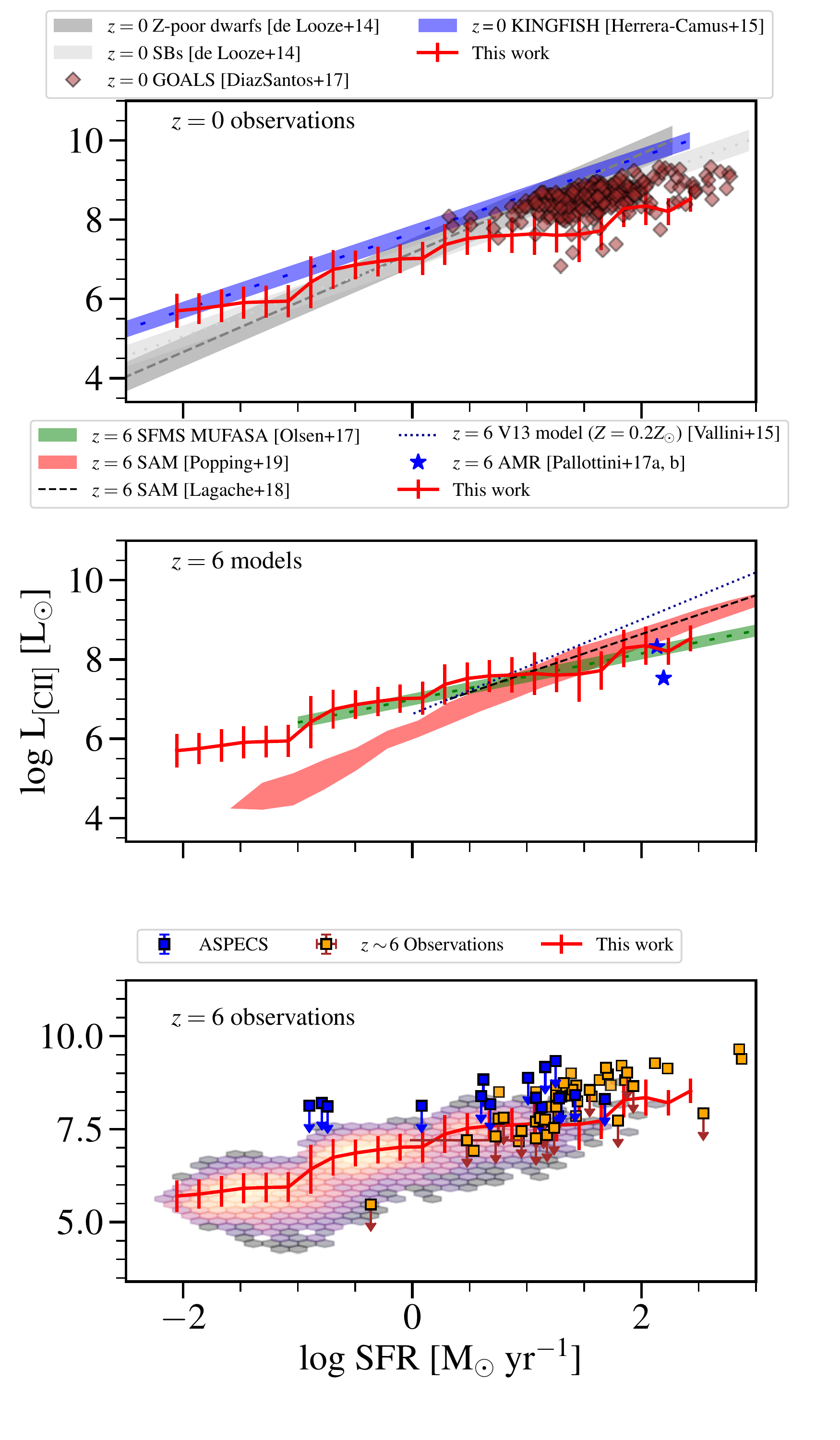}
\caption{Same as \Fig{ciisfr}, but visualized across three panels for clarity.
Top: Running mean and standard deviations of $z$\eq6 \simba galaxies (red) overplotted with $z$\eq0 \obs. The predicted \Lcii/SFR for galaxies with SFR $\gtrsim 1$ are 1-2 dex lower than observed galaxies in the local Universe, and the slope of the \Lcii vs. SFR relation is shallower.
Middle: \simba galaxies (red line) overplotted with other $z$\eq6 models in the literature (see legend). Our model predictions are in reasonable agreement with those of other studies, especially at higher SFR, but we predict a shallower \Lcii vs. SFR relation than other studies.
Bottom: \simba galaxies (red line and hexbin) overplotted with $z$\eq6 \obs. The hexbins are color-coded by the density
of points, see \Fig{ciisfr} for colorbar. Our predictions show reasonable overlap with the locus of the heterogeneous observational samples, although we do not produce any galaxies with \Lcii values as high as those of some of the detected galaxies at high SFR $\gtrsim 10$. See text for a discussion of possible reasons for these discrepancies.
\label{fig:ciisfrmultipanel}}
\end{figure}

In \Fig{ciisfr}, we plot the simulated \Lcii and SFR\footnote{The SFR is computed by dividing the stellar mass formed over the past 100\,Myr by this timescale.} of the \simba galaxies
together with measurements from existing observations at $z\,\simeq$\,6,
local measurements, and other model predictions at $z\,\simeq$\,6.
The \Lcii-SFR relation converges across the different simulation volumes as does the \Lcii-$M_{\rm mol}$ relation as seen in the right panel Fig.\,\ref{fig:mH2} in the Appendix.
For clarity, information shown in this figure is also plotted across three panels in \Fig{ciisfrmultipanel}.
We fit a linear model to \Lcii and SFR in log-log space to facilitate comparison with literature work, and obtain
\begin{equation}
\log L_{\rm [CII]} = (6.82\pm0.08) + (0.66\pm0.01)\times\log {\rm SFR},
\label{eqn:ciisfr}
\end{equation}
where \Lcii is in units of \Lsun, and SFR is in units of \Msun\,yr\pmOne.

The \cii luminosities of the \simba galaxies are consistent with existing upper limits and a handful of detections
 from targeted observations \citep[e.g.,][]{Ouchi13a, Kanekar13a, Ota14a,
Gonzalez-Lopez14a, Maiolino15a, Schaerer15a, Capak15a, Willott15a, Inoue16a, Pentericci16a,
Knudsen16a, Inoue16a, Bradac17a, Knudsen17a, Decarli17a, Smit17a, Carniani18a}.
In addition, our results are in agreement with the latest upper limits placed at $z\simeq\,$6 by the ALMA large program
ASPECS, which is an untargeted survey, placing an upper limit of
\Lcii$<$\,2\E8\,\Lsun for galaxies with UV-derived SFR of $\sim$\,0.25--50\,\Msun\,yr\pmOne (\citealt{Walter16a}, Uzgil et al., in prep.)\footnote{ASPECS consists of two bands (Bands 3 and 6). In Band 6, blind spectral scans over an 85 pointing mosaic
in the {\it Hubble} Ultra Deep Field (HUDF) with an areal footprint of 4.2 arcmin$^2$ were observed, reaching down to $\sigma_{\rm cont}$\eq9.3\,$\mu$Jy\,\bmm for the continuum and $\sigma_{\rm ch}$\eq0.3\,mJy\,\bmm per $\Delta v$\eq75\,\kms channel
for the line cube. At $z$\eq6, the sensitivity reaches a 5\,$\sigma$ limit of
 \Lcii$\simeq$\,10$^{8.3}$\,\Lsun at \z\eq6, assuming a linewidth of $\Delta v$\eq200\,\kms.\label{footnote:aspecs}}.
In particular, the ASPECS sources with upper limits on \Lcii shown in \Fig{ciisfr} with blue squares are
a combination of Lyman-$\alpha$ emitters (LAEs)
 with spectroscopic redshifts from the MUSE survey \citep{Inami17a}, 
 and Lyman break galaxies (LBGs)  \citep{Bouwens15a}.
 The Lyman-$\alpha$ luminosities of the LAEs are L$_{Ly\alpha}$\eq0.7\,$-$\,1.5\E{42}\,erg\,s\pmOne, with
 UV-based SFR\,$<$\,4\,\Msun\,yr\pmOne, and stellar mass of $\log(M_*/M_\odot)$\eq8.04\,--\,8.75
 (note that only two of the six MUSE LAEs have stellar mass constraints).
 The LBGs have H-band magnitudes of H$_{\rm 160}$\eq27.5--30.9\,mag, corresponding to a UV-based
 SFR of 0.25\,--\,48\,\Msun\,yr\pmOne, and have stellar masses between $\log(M_*/M_\odot)$\eq7.99\,--\,9.37.
In general, the running mean in \cii luminosity of the \simba sample is lower than the average of existing detections of {\em targeted} \obs at $z$\ssim6.

Our results are consistent with those based on the \ncode{Serra} simulation suite by \citet{Pallottini17a} and \citet{Pallottini17b},
which is a suite of cosmological zoom-in AMR simulations that resolve the gas down to
10\,pc-scales at $z\simeq$\,6.
It is also in reasonably good agreement with the sample of 30 \ncode{mufasa} galaxies
analyzed in \citet{Olsen17a}, thus broadly confirming these results
using a larger sample from its successor simulation (\ncode{simba}),
while reaching comparable resolution over cosmological volumes.
That said, our results yield a flatter \Lcii--SFR relation than other models at $z\,\simeq$\,6,
such as those based on SAMs and semi-empirical models \citep{Vallini15a, Lagache18a, Popping19a}.
We discuss the potential causes of the differences seen between our results and other models in the literature in \Sec{caveats}.

\subsection{ \cii Luminosity Function at \z$\simeq$\,6}
\begin{figure}[phtb]
\centering
\includegraphics[trim=0 0 0 0, clip, width=.5\textwidth]{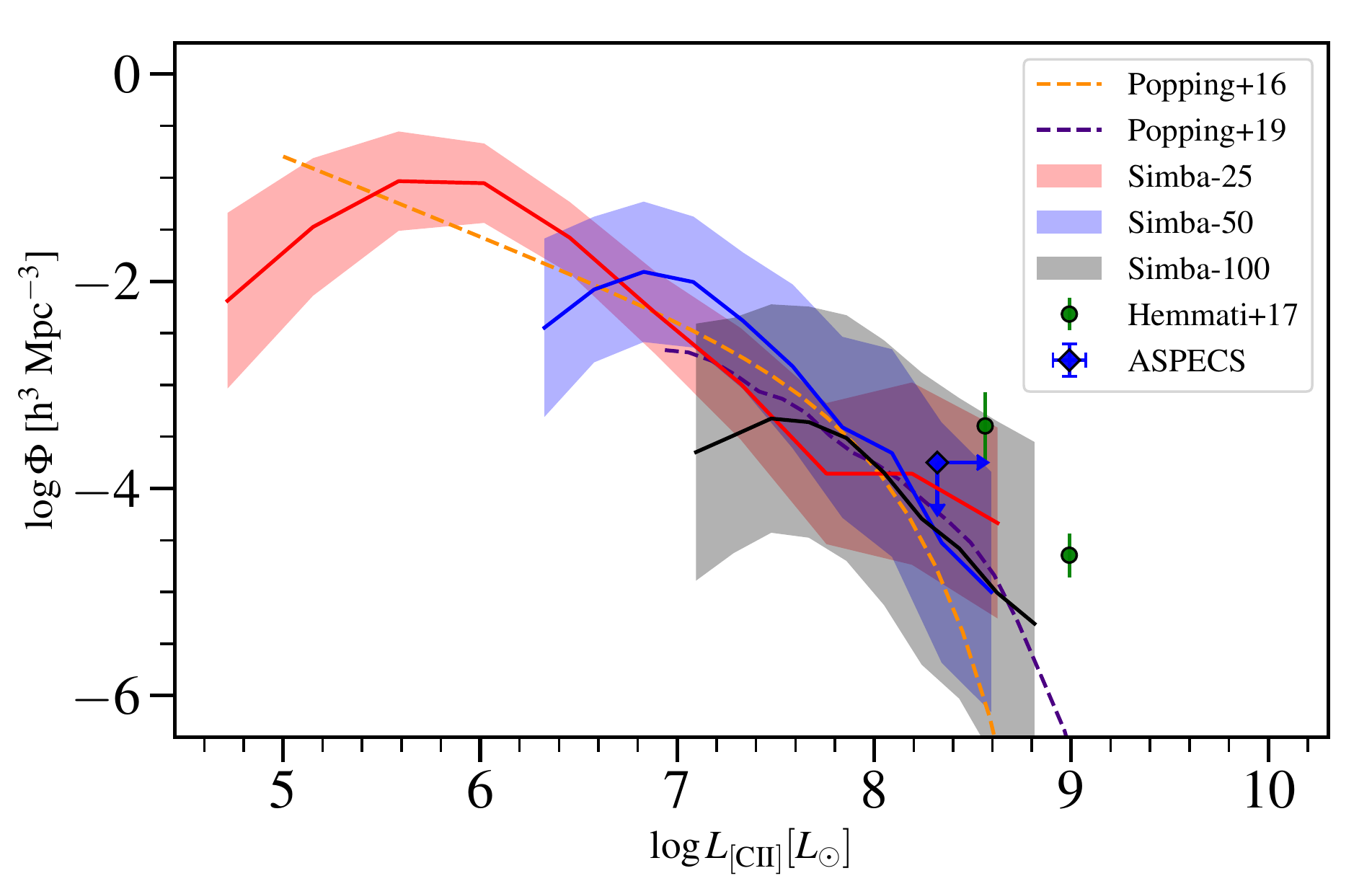} 
\caption{\cii LF predicted at \z$\simeq$\,6 based on the cosmological hydrodynamics simulation \simba.
Shaded regions are obtained by jackknife resampling of the simulation sub-volumes.
The flattening and turnover at the faintest end is due to incompleteness of
haloes with $\log$ \Lcii$\lesssim$\,6\,\Lsun.
Results from SAM-based models are overplotted as dashed lines \citep{Popping16a, Popping19a}.
Our results are fully consistent with the SAM-based model predictions within the error bars and the
upper limits from ASPECS (blue symbol; Uzgil et al., in prep.).
\label{fig:ciilf}}
\end{figure}

In \Fig{ciilf}, we show predictions for the \cii LF at \z$\simeq\,$6
based on the simulated \Lcii of galaxies in Simba-25, Simba-50, and Simba-100.
We note that previously, LF predictions were only possible in models that made
more simplified assumptions to connect \Lcii to dark matter halos.
Nonetheless, our results are in agreement with those based on SAMs by \citet{Lagache18a} and \citet{Popping19a},
and with constraints from the latest limits placed using data from ASPECS (Uzgil et al., in prep.; see footnote~\ref{footnote:aspecs}).

Using the \citet{Capak15a} targeted sample, \citet{Hemmati17a}
report a volume density that is almost an order of magnitude higher
than our results at the bright end at $\log $ \Lcii$\gtrsim$\,8.5\,\Lsun, although they
are consistent within the error bars. Note that the actual uncertainties on the \cii LF constrained by
the \citet{Capak15a} sample are likely to be larger than those reported by \citet{Hemmati17a}, as
incompleteness and selection bias are not corrected for.
As shown in \Fig{ciisfr}, most of the \simba galaxies have \Lcii\eq10$^{6-8}$\,\Lsun.
Thus, it is unsurprising to see a discrepancy in the \cii LF between the \citet{Capak15a} sample and our sample
due to the lack of overlap in terms of \Lcii.

\citet{Miller16a} derive a \cii LF using the Bolshoi-Planck dark matter only simulation catalog from
\citet{Behroozi13b}, the abundance matched SFR from \citet{Hayward13c},
and the empirical \Lcii--SFR relations established at $z$\ssim0 by \citet{DeLooze14a}.
Consistent with our results, \citet{Miller16a} report a \cii LF that underpredicts the observational constraints placed
by \citet{Hemmati17a} using the \citet{Capak15a} sample and that placed based on a
blind search of five deep fields centered on IR-bright galaxies and quasar host galaxies at $z$\ssim6.
By simulating only regions with \Lcii matched to the central galaxies observed in
the deep fields (8.7\,$<\,\log$\Lcii/\Lsun$<$\,9), \citet{Miller16a} find a good agreement between the \cii LF
and the observational constraints.
On this basis, they argue that the \cii-detected sources in the deep fields are indeed in biased overdense regions.
This may partially explain the discrepancy seen between the observed and the \simba based \cii LF ---
since the largest simulation box of \simba is 100\,cMpc, and it may not contain these rare highly biased regions.

\subsection{\Lcii\,--\,M$_{\rm halo}$ Relation} \label{sec:halo}
Forecasts for upcoming \cii LIM surveys have been obtained using
scaling relations between \cii luminosity and halo mass, where
the latter quantity is obtained from large volume N-body simulations (e.g., \citealt{Silva15a, Kovetz17a}).
The large cosmological volumes provided by N-body simulations are needed to make mock lightcones
for LIM \citep[e.g.][]{yang2020}; however, 
most previous work in this area has made simple empirical assumptions regarding the relationship between
dark matter halo properties and \cii line luminosity.
Here we study the \Lcii\,--\,M$_{\rm halo}$ relation based on
the central galaxies in our cosmological hydrodynamic simulation.

\begin{figure}[htbp]
\centering
\includegraphics[trim=0 0 0 0, clip, width=.5\textwidth]{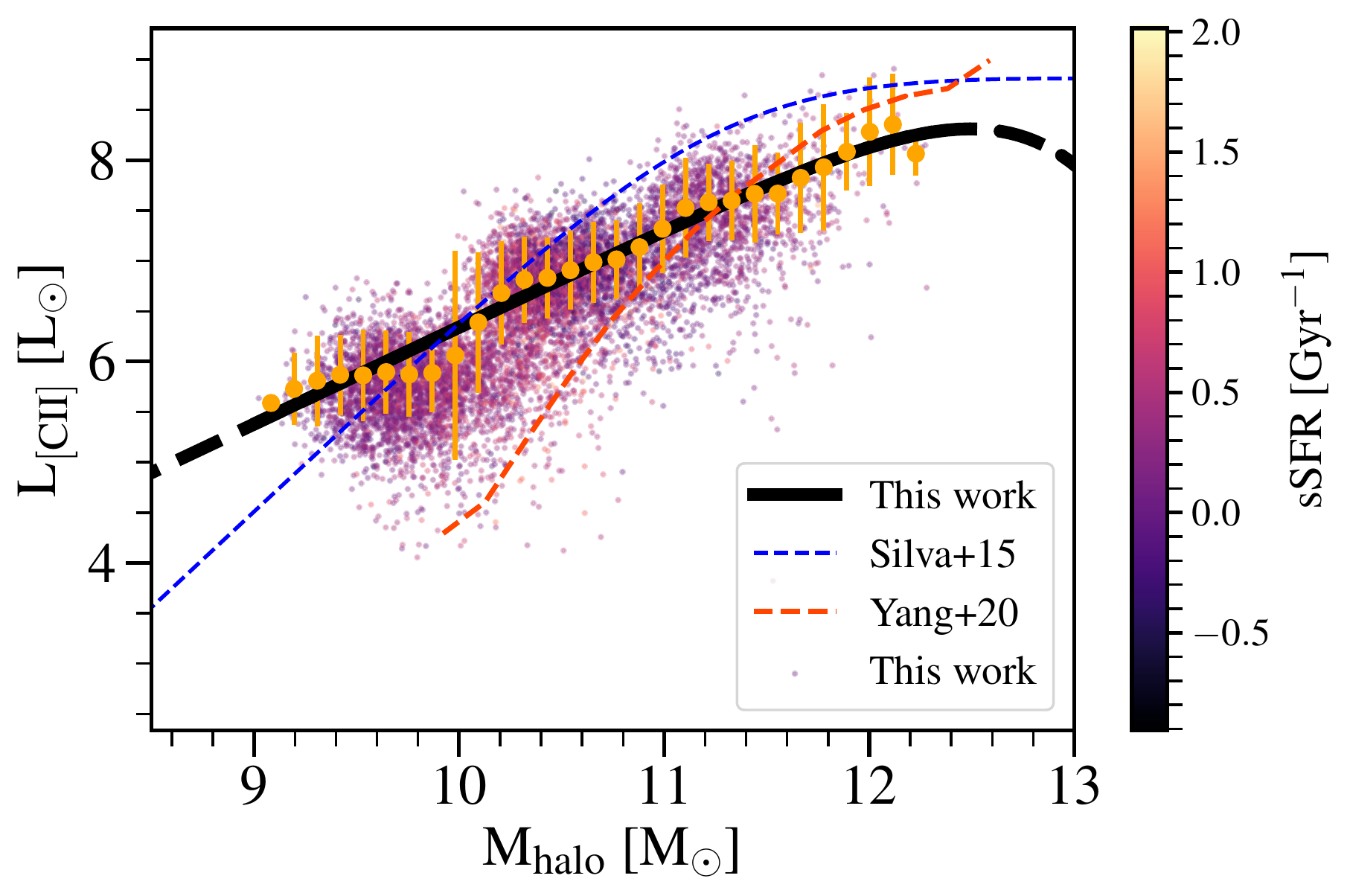}
\caption{\Lcii$-$\,M$_{\rm halo}$ of \simba galaxies studied in this work, color-coded by sSFR (dot symbols).
The black line shows the best-fit parametric model, whereas the orange dots show
the mean 
when the \simba data is binned in 30 logarithmic intervals in M$_{\rm halo}$ (i.e., non-parametric).
The blue dashed line shows the m1 model of \citet{Silva15a} at a comparable redshift.
The red line shows the model from \cite{yang2020} based on SAMs by \citet{Popping19a}. Although there is qualitative agreement between the different models, the remaining discrepancies could have significant implications for predictions for upcoming line intensity mapping surveys.
\label{fig:ciimhalo}}
\end{figure}

In \Fig{ciimhalo}, \simba galaxies are shown in \Lcii versus M$_{\rm halo}$ with a fit made following the formalism of \citet{Silva15a}, where SFR is expressed in terms of M$_{\rm halo}$:
 \begin{equation}
{\rm SFR} = M_0 \times \left(\frac{M_{\rm halo}} {M_a} \right)^a \left(1 + \frac{M_{\rm halo}}{M_b}\right)^b.
\end{equation}
Together with the expected linear relation between \Lcii and SFR, the following equation relates M$_{\rm halo}$ to \Lcii:
\begin{equation}
\log L_{\rm [CII]} = M'_0 + a' \log \left(\frac{M_{\rm halo}} {M'_a} \right) + b' \log\left(1 + \frac{M_{\rm halo}}{M'_b}\right),
\label{eqn:lciisfr}
\end{equation}
where \Lcii is in units of \Lsun, and M$_{\rm halo}$ is in units of \Msun. With the current set of parameters adopted in the sub-grid model (see \Sec{sigame}), the fitted parameters are
$a'$\eq0.65,
$b'$\eq$-$9.85,
$M'_0$\eq3.62,
$M'_a$\eq1.50\E{7}, and
$M'_b$\eq2.90\E{13}.

The \Lcii--\,M$_{\rm halo}$ relation of central galaxies in \simba has
a scatter of $\simeq$\,0.5\,dex around the fit.
We also show the running mean in the same figure which follows the parametric form.
In the same figure, we show a comparison with the \Lcii--\,M$_{\rm halo}$ relation from
a SAM \citep{Popping19a,yang2020}, which is steeper than both \citet{Silva15a} and
this work; but is however, consistent within the scatter of \simba galaxies.

\section{Discussion: Discrepancies and Caveats} \label{sec:caveats}

\begin{turnpage}
\input{Table_models}
\end{turnpage}

As discussed in \Sec{results}, our predicted \cii LF is consistent
with that predicted from other models based on SAMs, and the \simba
galaxies lie in the same region of \Lcii--\,SFR as the galaxies
studied by one of the most detailed cosmological zoom-in AMR
simulations at the same redshift \citep{Pallottini17a, Pallottini17b}.
However, compared to other models, our model underpredicts the \cii
luminosity at the bright end (and high SFR; \Fig{ciisfrmultipanel})
and yields a flatter \Lcii--\,SFR relation.  This discrepancy may
arise from for instance different ranges of properties for galaxies in
the samples, different predicted scaling relations between galaxy
properties in different galaxy evolution models, a limited number of
massive halos in our simulation, and the different sub-grid treatments
of physical processes in \simba and in \sigame compared to other
approaches adopted in the literature used for comparison here
(\Tab{model}).  For instance, \citet{Vallini15a} simulate the \cii
line emission by post-processing the UV radiation field of an
SPH-based simulated galaxy using \ncode{Licorice}, and calculate the
\cii emission using a combination of an analytical model and the
photo-dissociation region (PDR) code \ncode{ucl\_pdr}
(\citealt{Bayet09b} and references therein; to account for
contributions from PDR).  While the sub-grid modeling of
\citet{Popping19a} follows a similar approach as \sigame, the former
is based on a SAM while the latter is applied to hydrodynamical
simulations.  As a result, there are relatively subtle differences
between the two, such as the assumption of exponential gas disks for
all galaxies in the former.  In addition, the former approach assumes
that all GMCs in each galaxy share the same metallicity based on the
global metallicity, and adopts \ncode{despotic} instead of
\ncode{cloudy} in performing the thermochemistry calculation. As
mentioned in \Sec{introduction}, \ncode{despotic} does not account for
line emission in the ionized phase while the latter does.  In contrast
to \citet{Popping19a}, \citet{Lagache18a} use \ncode{cloudy} to
post-process their SAM. While using \ncode{cloudy} is the same
approach as \sigame (including this work) and works by e.g.,
\citet{Katz19a} and \citet{Pallottini19a}, \citet{Lagache18a} adopt
different sub-grid approaches and assumptions compared to those made
for hydrodynamical simulations.  In particular, their model does not
account for \cii emission coming from regions outside of PDRs, the
complexity of the multiphase ISM, and the detailed structures of
molecular clouds.  This illustrates the various differences between
existing models which can contribute to the discrepant \Lcii--\,SFR
slopes.  Using SAMs, \citet{Popping19a} experiment with different
assumptions made in the sub-grid approaches and indeed find
differences in the resulting [CII] luminosity.

\begin{figure}[htbp]
\centering
\includegraphics[trim=0 0 0 0, clip, width=.5\textwidth]{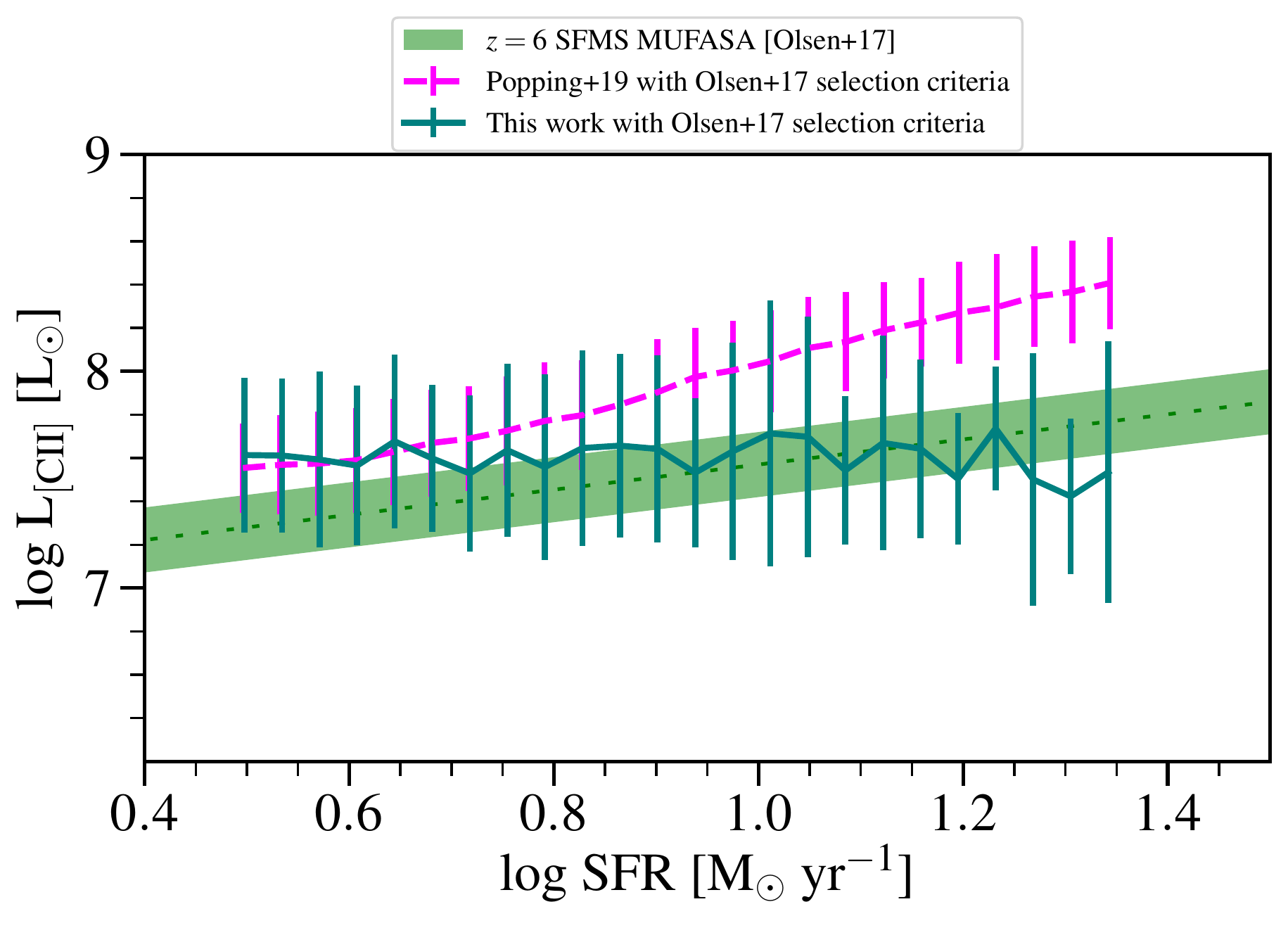}
\caption{
Region of \Lcii--\,SFR in log-space spanned by \simba galaxies (teal line) and
galaxies from Popping et al.~(\citeyear{Popping19a}; magenta line)
selected with the same range in stellar mass, SFR, and gas-phase metallicities
as \citet{Olsen17a} (green shaded). A relation with a flatter slope than other models persists for the \simba galaxies (see \Sec{caveats}).
\label{fig:ciisfr_selectedolsen17}}
\end{figure}

\citet{Lagache18a} report that after selecting galaxies with the same range of stellar mass, SFR, and gas-phase metallicities
as the \ncode{Mufasa} sample studied by \citet{Olsen17a} --- i.e.,
with $M_*\in$\,(0.7--8)\E{9}\,\Msun, SFR$\in[$3--23$]$\,\Msun\,yr\pmOne, and $Z_{\rm gas}\in[$0.15--0.45$]$\,$Z_{\odot}$ --- the \Lcii--\,SFR relation of their SAM galaxies is flatter and
more consistent with that of \citet{Olsen17a}.
Yet, this relation remains flatter than that obtained when applying the same set of criteria to the
SAM galaxies of \citet{Popping19a} (9,653 galaxies after selection).
In \Fig{ciisfr_selectedolsen17}, we show the \Lcii --\,SFR for
a subset of \simba galaxies selected based on these criteria (1,136 galaxies)
compared to \citet{Olsen17a} and \citet{Popping19a}.
A relation with a flatter slope than \citet{Popping19a}'s SAM-based models persists for the \simba galaxies.

\begin{figure}[htbp]
\centering
\includegraphics[trim=0 0 0 0, clip, width=.5\textwidth]{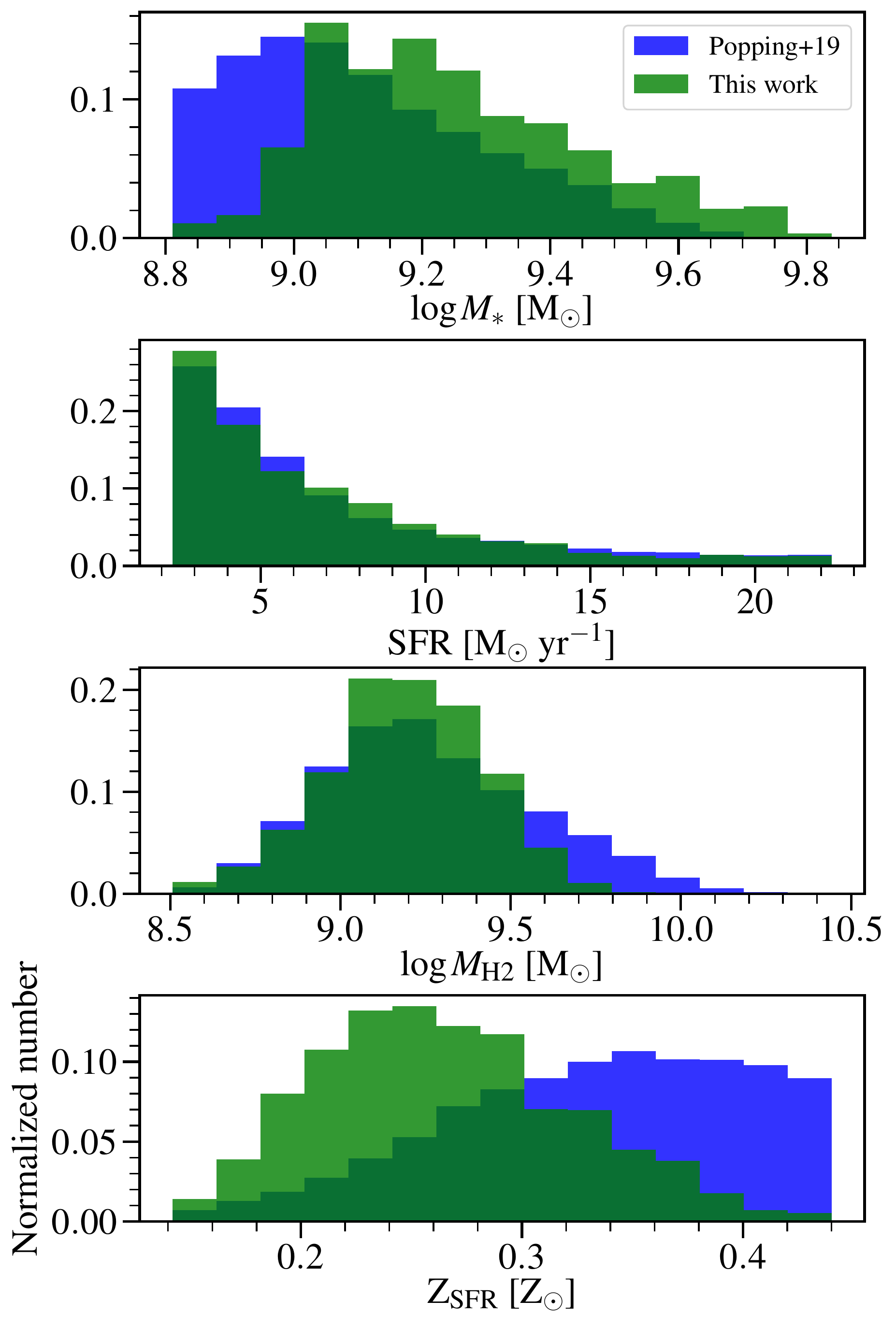}
\caption{Normalized distributions of stellar mass, SFR, molecular gas mass, and metallicity between the subset of galaxies
from \citet{Popping19a} (blue) and this work (green) after applying the \citet{Olsen17a} selection.
While distributions in SFR are comparable, the subset of galaxies from the \citet{Popping19a} sample have higher
molecular gas mass and metallicity compared to the \simba subset.
\label{fig:p19thiswork_dist_selectedolsen17}}
\end{figure}

\begin{figure}[htbp]
\centering
\includegraphics[trim=10 150 30 100, clip, width=.5\textwidth]{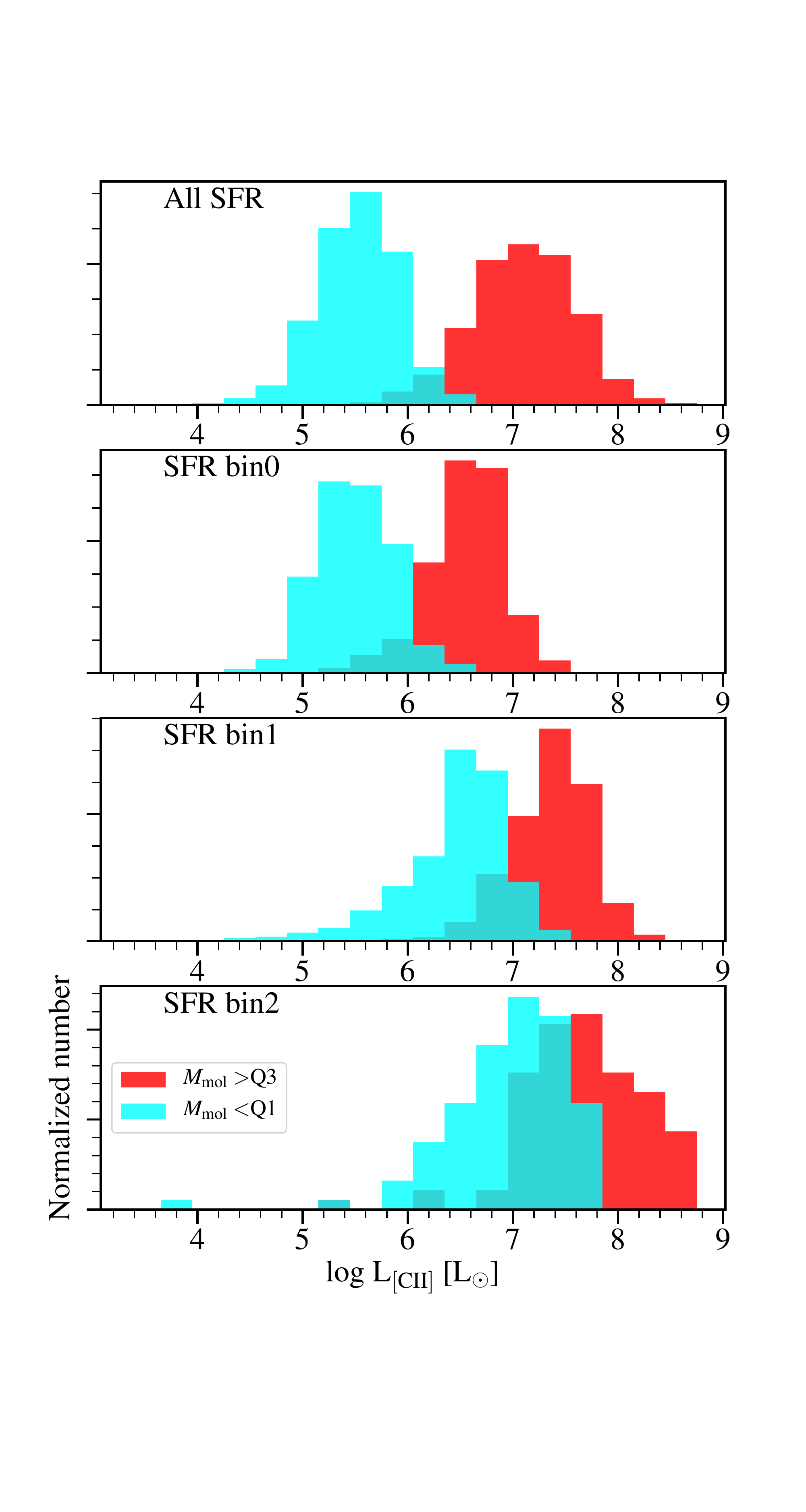}
\caption{
Distributions of \Lcii across all \simba galaxies, in the first (red) and third (cyan) quartiles in molecular gas mass
from different SFR bins (second through last panels). See text for the SFR covered by each bin.
The top panel shows the distributions across all SFR.
At a fixed SFR, galaxies with higher $M_{\rm mol}$ have higher \Lcii.
\label{fig:ciisfr_quartile_mol}}
\end{figure}

\begin{figure}[htbp]
\centering
\includegraphics[trim=10 150 30 100, clip, width=.5\textwidth]{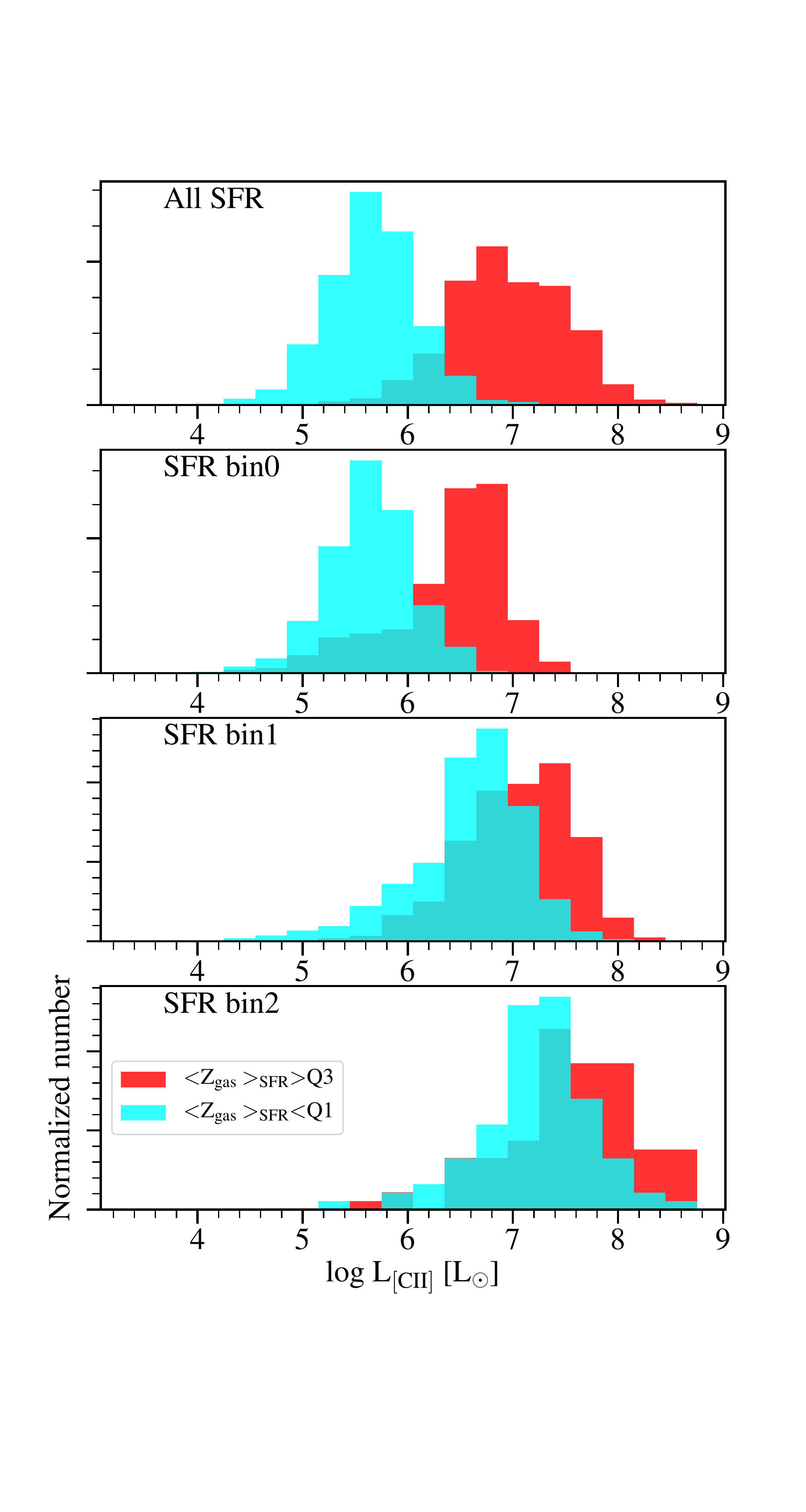}
\caption{
Same as \Fig{ciisfr_quartile_mol} but for metallicity.
At a fixed SFR, galaxies with higher $\langle Z_{\rm gas}\rangle_{\rm SFR}$ have higher \Lcii.
\label{fig:ciisfr_quartile_Zgas}}
\end{figure}

\begin{figure}[htbp]
\centering
\includegraphics[trim=10 150 30 100, clip, width=.5\textwidth]{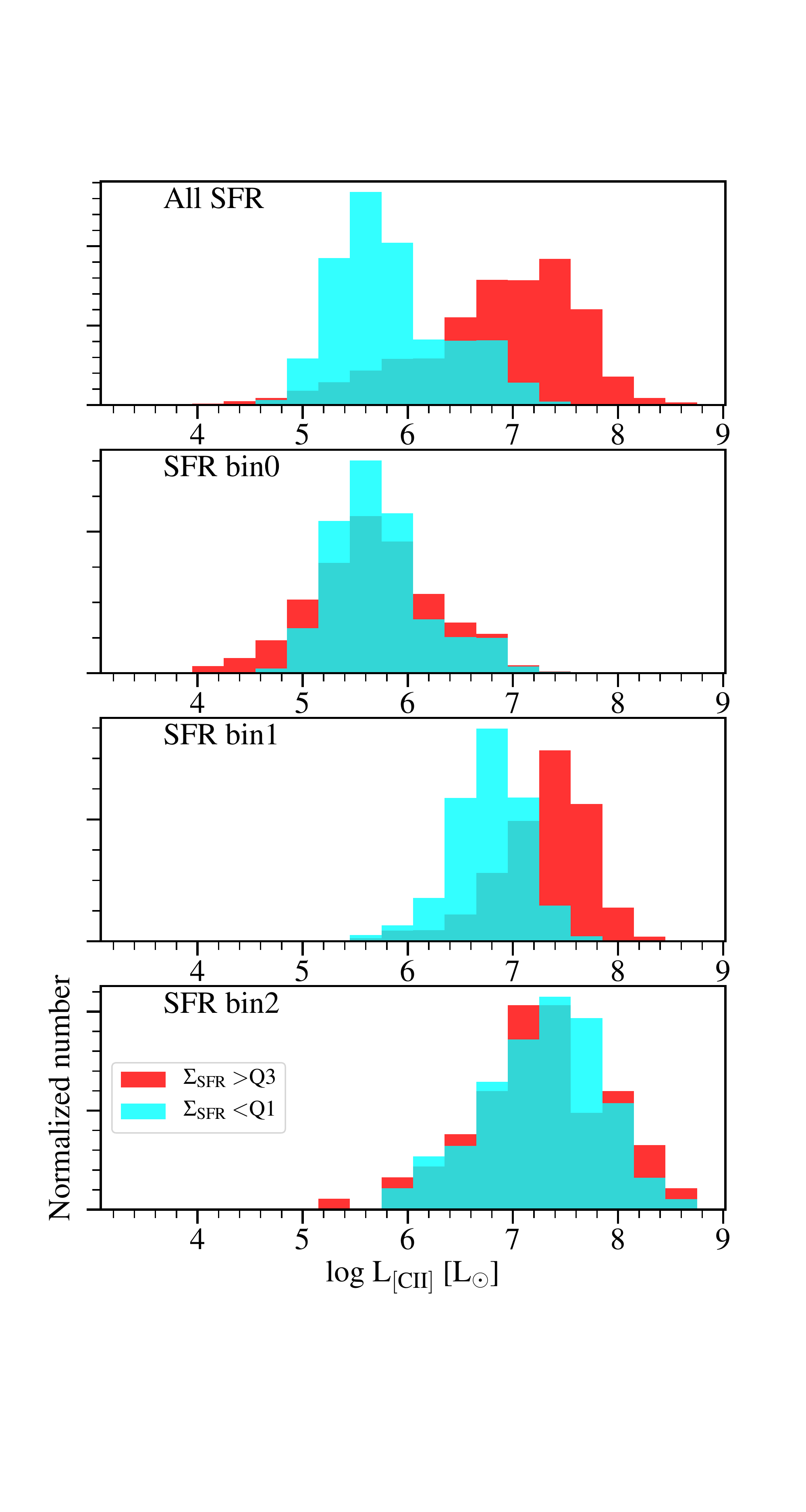}
\caption{
Same as \Fig{ciisfr_quartile_mol} but for SFR surface density.
Overall, galaxies with higher $\Sigma_{\rm SFR}$ have higher \Lcii, but this trend is not seen after accounting for the influence caused by different SFR.
\label{fig:ciisfr_quartile_sfrsd}}
\end{figure}

As shown and discussed in the literature (e.g., \citealt{Vallini15a, Olsen17a, Lagache18a, Pallottini19a, Popping19a}),
a lower metallicity can result in a lower \Lcii at given SFR.
In fact, as shown in \Fig{p19thiswork_dist_selectedolsen17}, while the distributions in SFR
between the subset of galaxies from the \citet{Popping19a} sample and this work are comparable
after applying the \citet{Olsen17a} selection cut, the former are more metal rich, with higher molecular gas masses
compared to the latter.
The trend of decreasing molecular gas mass and metallicity with \Lcii is also seen in Figures~\ref{fig:ciisfr_quartile_mol} and \ref{fig:ciisfr_quartile_Zgas},
where we show the different \Lcii distributions when the full set of 11,137 \simba galaxies are selected
using the first and third quartiles in molecular gas mass and metallicity, respectively.
To account for the effect of SFR in \Lcii, we bin the galaxies into three SFR bins, $\in[0.01, 10.3]$, $\in[0.3, 10.3]$, $\in[10.3, 329]$\,\Msun\,yr\pmOne\footnote{Binning in SFR is performed in log-space to avoid small number statistics in each bin and to ensure that the number of galaxies in each bin is of the same order of magnitude (see e.g., hexbins in \Fig{ciisfr}).}.
At least in the lowest SFR bin, the variation seen in the \cii luminosity is mostly driven by the lower metallicity and molecular gas mass (see also \citealt{Narayanan17a}).
Across all SFR, galaxies with lower \Lcii correspond to those with the least molecular gas mass, metallicity,
and SFR surface density (Figures~\ref{fig:ciisfr_quartile_mol}, \ref{fig:ciisfr_quartile_Zgas}, and \ref{fig:ciisfr_quartile_sfrsd}).

As mentioned in \Sec{introduction}, models in the literature adopt different approaches and assumptions in post-processing the simulations. This could also yield different simulated line luminosities.
 A steeper slope, more compatible with other models in the literature, namely \citet{Vallini15a} and
 \citet{Popping19a},
 would result from adopting the local ISM abundance ratios\footnote{The C/H ratio of
 the local ISM is comparable to the Solar value \citep{Cowie86a, AllendePrieto02a, Asplund09a}.}
 instead of the abundance pattern tracked in \simba (see \Sec{element} and \citealt{Olsen17a}).
Both \citet{Vallini15a} and \citet{Popping19a} adopt a Solar abundance pattern (of relevance to this work is the C/H ratio)
and scale the C/H ratio according to the gas-phase metallicity of each galaxy to determine the \cii emissivity.
That is, their models do not consider abundance patterns that differ from Solar.
The resulting \Lcii could differ significantly owing to the amount of cooling via C$^+$.
On the other hand, \citet{Lagache18a} accounted for abundance ratios that differ from Solar. Specifically, they
scale the element abundances based on the median of their sample. Unsurprisingly,
the slopes of the \Lcii--\,SFR relation of \ncode{mufasa} and \simba galaxies
are the most compatible to the \citet{Lagache18a} model.
As mentioned by \citet{Lagache18a} and \citet{Katz19a},
details in modeling the interstellar radiation field intensity, self-shielding,
and the choice and implementation of stellar feedback are all effects that can
cause differences between existing models (see also \citealt{Pallottini17a}).

In contrast to \ncode{mufasa}, where dust is not tracked in the simulation,
causing \citet{Olsen17a} to adopt a constant DTM ratio,
\simba tracks dust in the simulation; however, we do not create a different \ncode{cloudy} lookup table for
each DTM ratio since it would become computationally intractable.
The mean DTM of \simba galaxies is $\xi_{\rm DTM}$\eq0.25, which yields a \cii luminosity approximately
$\lesssim$\,0.5\,dex higher than for models with a DTM ratio set to the Solar value of
$\xi_{\rm DTM}$\eq0.46 (see Appendix; cf. \citealt{Olsen17a}  who finds that \Lcii only increased by $\sim0.15$ \,dex at a given SFR when decreasing the DTM by a factor 2 from Solar DTM for their \ncode{Mufasa} SFMS sample\footnote{This work uses
the first release of \sigame; see main improvements in \sigame in \Sec{sigame}.}).
Note that in reality, the DTM ratio varies from galaxy to galaxy (and within galaxies themselves), and is correlated with
the galaxy metallicity \citep[e.g.,][]{Remy-Ruyer14a, Popping17a, DeVis19a}.
Since the standard deviation of the DTM ratio of the \simba sample studied here is $\sigma$\ssim0.15,
we do not expect the simulated \cii luminosity to deviate more than 0.5\,dex as a result of
variations in the DTM ratio.

A final important caveat is that the effect of AGN was not included in the present modeling with \ncode{cloudy}, although \ncode{cloudy} does have the capability to do so and this has been shown to be relevant at least for CO line emission at high redshift \citep{Vallini19}, and hence likely also relevant for [CII]. The effect of AGN is an additional feature that we wish to include in the future.

\section{Summary and Conclusions}      \label{sec:conclusion}

In this work, we presented the first prediction of the \cii luminosity
function (LF) during the Epoch of Reionization (EoR; $z\simeq$\,6) based
on large-volume cosmological hydrodynamical simulations (\simba)
coupled with radiative transfer and line spectral synthesis
calculations.  We simulate the \cii line luminosity for a sample of
11,137 galaxies identified in the combined 25, 50, and
100\,\cmpc\ boxes (with 2\,$\times$\,1024$^3$ particles each) at $z\simeq$\,6.  The runs for the three simulation
boxes have identical input physics,  and
produce converged GSMF and SFR functions without fine-tuning
the parameters in the sub-grid models of \simba.  In addition, both
GSMF and SFR functions are in good agreement with observations at this redshift.  This
is crucial as we make predictions for the \cii LF in the luminosity
range of 5.5\,$<$\,$\log$(\Lcii/\Lsun)$<$\,8.5 by combining the boxes.

We use an updated version of \sigame to post-process the \simba output. Three major improvements are
implemented relative to the previous version of \sigame presented in \citet{Olsen17a} to produce the results presented;
{\em (i)} We do not fix the number and width of shells for each GMC model, but instead allow \ncode{cloudy} to determine the optimal quantities to ensure convergence.
This modification leads to more accurate calculation of the grain photoelectric heating of the gas and
increases the importance of gas heating due to this mechanism in the GMC models --- the main excitation mode for \cii emission;
{\em (ii)} The number of \ncode{cloudy} models used to create look-up tables is significantly increased
from 1296 to 4096 models to ensure better sampling of the physical properties of the ISM,
and
{\em (iii)} A DTM ratio of 0.25 is adopted based on the mean value of the \simba galaxy sample
rather than using a solar DTM value.
Finally, \citet{Olsen17a} simulated line emission for a subset of 30 zoom-in \ncode{Mufasa} galaxies along
the SFMS, whereas with \simba, we are able to expand the parameter space in M$_{\rm halo}$, SFR, M$_*$, M$_{\rm gas}$, SFR surface density, and metallicity at comparable resolution without the need for ``zooming in'' on specific galaxies.

We summarize the main results of this paper in the following:
\begin{itemize}
\item The simulated \Lcii is consistent with the range observed in \z\ssim6 galaxies, with
a spread of $\simeq$\,0.3\,dex at the high SFR end of $>$\,100\,\Msun\,yr\pmOne
which increases to $\simeq$0.6\,dex at the lower end of the SFR. The predicted \Lcii-SFR is consistent with targeted observed samples within the uncertainties due to selection and incompleteness effects. On the other hand, our model does not produce galaxies with values of \Lcii as high as those for some galaxies observed
in targeted heterogenous samples reported in the literature, at a given SFR.

\item The \cii LF is consistent with the upper limits placed by the only existing
untargeted flux-limited \cii survey at the EoR (ASPECS) and those predicted by semi-analytic models.

\item Our model yields a \Lcii--\,SFR relation similar to \ncode{mufasa} \citep{Olsen17a} but
is flatter compared to some other models in the literature. The flatter slope results from
different galaxy properties sampled by different simulations, implementation and assumptions of sub-grid recipes between different simulations and the post-processing steps (see \Tab{model}).

\item At a fixed SFR, galaxies with higher molecular gas mass, metallicity, and SFR surface density have higher \cii luminosity.

\item We present the \Lcii--\,M$_{\rm halo}$ relation for the central galaxies in
\simba at $z\sim6$.
Our relation is steeper than those based on N-body simulations and SAMs; but are consistent within the scatter of $\simeq$\,0.5--0.6\,dex.

\end{itemize}

The differing results presented in the literature on simulating \cii
line emission at the EoR highlights the challenges modelers face in
this field.  As discussed in this paper, SAMs are computationally
efficient and can simulate the line emission for a statistically
significant sample of galaxies, but SAMs have their limitations. They
do not contain information regarding the structure of the ISM of
galaxies, or the 3D distribution and morphology of galaxies, to name a
few. Cosmological hydrodynamic simulations, on the other hand, can
provide more detailed information on the temperature and density of
the intergalactic medium and interstellar medium, the 3D structures of
galaxies, and the local properties of galaxies (e.g., each gas element
has a different metallicity), but are computationally demanding. In
addition, large volume cosmological simulations still lack the
resolution needed to resolve the multi-phase ISM in detail.  Our convergence tests among different resolution simulations indicate good convergence in stellar masses and star formation rates but less than ideal convergence in the molecular gas fractions, indicating that some sub-grid models may still need refinement to improve convergence properties. Zoom-in
simulations have been used to achieve higher resolution, but the
computational cost limits the number of galaxies (and thus, the galaxy
parameter space studied) that can be ``re-simulated'' with the zoom-in
approach in reasonable time.  Calibrating parameters based on higher
resolution simulations with resolved ISM properties will be necessary to test the
sub-grid models and assumptions made, for example adopting the
distributions based on pc-scale hydrodynamic simulations (e.g.,
\citealt{Tress20a}), and such work is underway by \sigame group members.

Deep galaxy surveys over the past decade have provided constraints on the cosmic star formation
history and supermassive black hole growth history \citep[e.g.,][]{Madau14a, Wilkins19a, Hickox18a, Aird19a},
while surveys and intensity mapping experiments planned in the next decade, with facilities such as
the {\it James Webb Space Telescope}, {\it WFIRST}, {\it Euclid}, LSST,
CONCERTO, HERA, SPHEREx, EXCLAIM, and TIM \citep[see reviews by][]{Kovetz17a, Cooray19a},
will expand and sharpen our view of galaxy evolution by discovering galaxies in new ways, obtaining photometric and spectroscopic redshifts, and
probing a wider parameter space in galaxy properties and large scale environment.
Due to the brightness of the fine-structure line \cii and its accessibility at high redshift,
it is one of the main spectral lines that may be observed at the EoR to study the ISM properties of galaxies and
to secure their spectroscopic redshifts. LIM experiments such as CONCERTO, TIME, and
CCAT-p will measure the \cii line power spectrum from galaxies at the EoR; however, the detection limits
and interpretation of the observed power spectrum depend on the \cii line luminosities of different EoR galaxy populations within
the volume sampled. The fainter populations are below the current detection limit of existing
facilities, but their signal can be predicted, highlighting the importance of building theoretical frameworks
to simulate the \cii line at the EoR using cosmological simulations.

\appendix

\section{H2 gas mass}
\begin{figure*}[htbp]
\centering
\includegraphics[trim=0 0 0 0, clip, width=.45\textwidth]{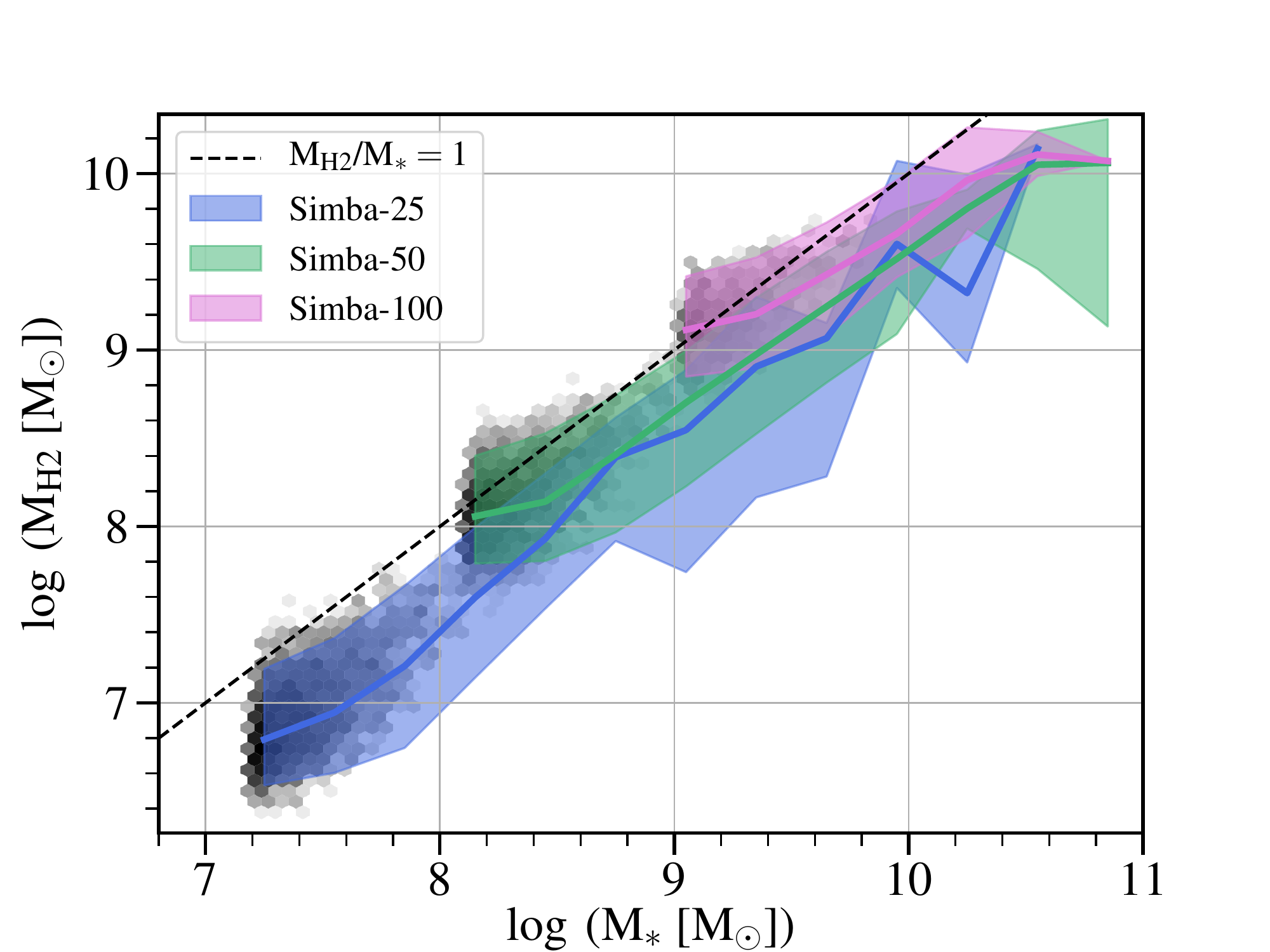}
\includegraphics[trim=0 0 0 0, clip, width=.45\textwidth]{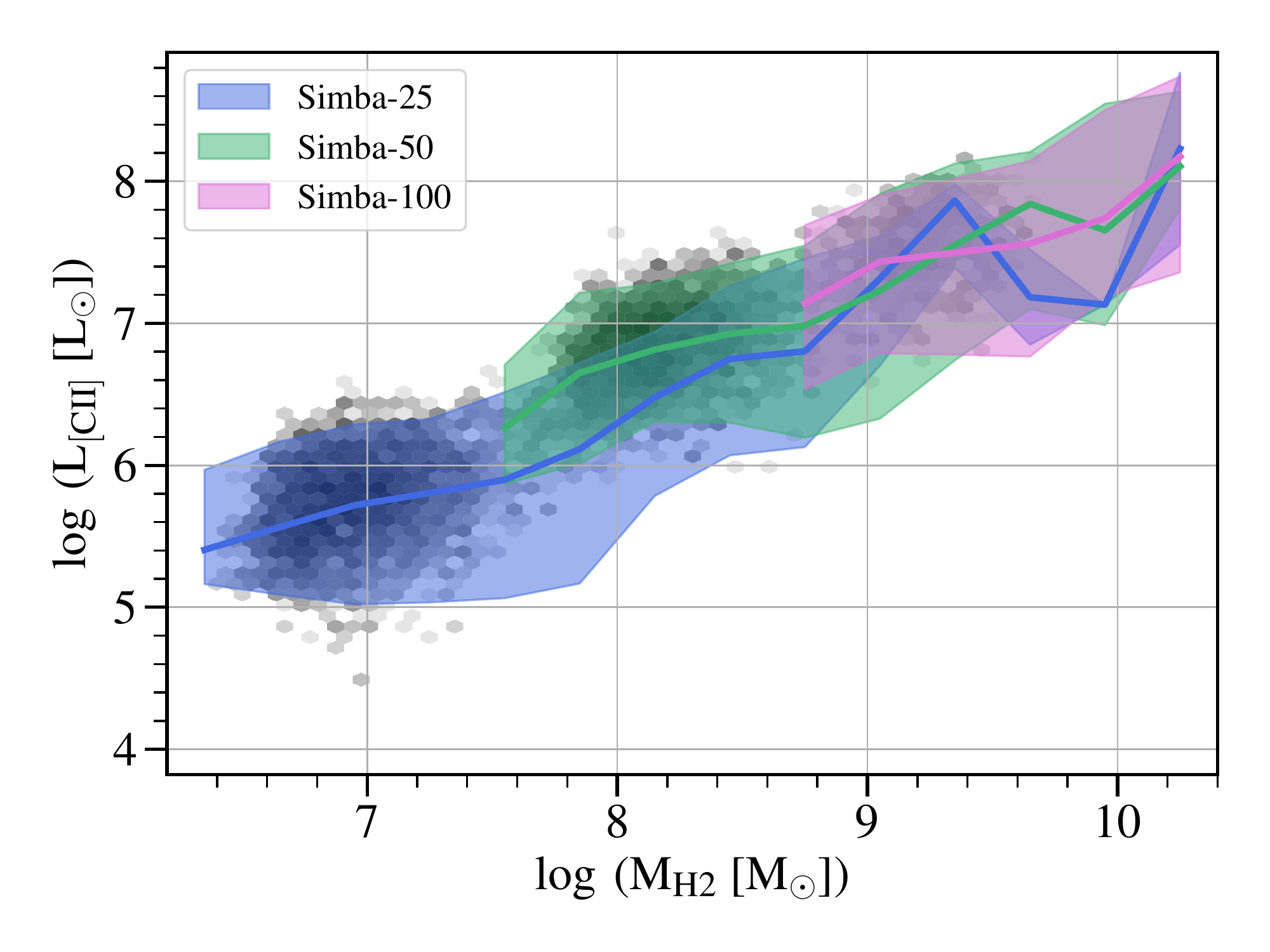}
\caption{{\it Left}: Molecular mass against stellar mass for the entire \simba galaxy sample (grey hexbin contours) and median molecular mass for each simulation box separately. Shaded areas indicate the 0.05 to 0.95 quantiles around the median values. A black dashed line indicates a 1-to-1 relation, showing how the molecular to stellar mass fraction generally increases with stellar mass for the \simba simulations. {\it Right}: \cii luminosity against molecular mass for all \simba volumes with the same plotting technique. Despite the increase in molecular to stellar mass fraction, all volumes agree with a common \cii molecular gas mass factor.
\label{fig:mH2}}
\end{figure*}

The left panel of Fig.\,\ref{fig:mH2} is an alternative way to show the information displayed in the middle panel of Fig.\,\ref{fig:GSMF}, by replacing the plotting method with hexbin contours and showing molecular gas mass on the y axis instead of molecular to stellar mass fraction. The median molecular gas mass is shown on top of the hexbin contours for each simulation box separately. In Fig.\,\ref{fig:GSMF}, the different simulation box sizes seem inconsistent with each other, but when looking at Fig.\,\ref{fig:mH2} the in-continuities largely go away, and we instead see a slightly increasing molecular gas mass fraction with stellar mass. Unfortunately we do not properly cover the stellar mass range $> 10^{10}\,$\Msun to compare with observations showing a slight downward trend of $M_{\rm mol} /M_{*}$ fractions with stellar mass in this high mass range \citep{Saintonge2011}. The right panel of Fig.\,\ref{fig:mH2} shows how the \cii luminosity per molecular gas mass corresponds to a common \cii molecular gas mass conversion factor across the different \simba volumes. The use of \Lcii as a gas mass tracer with this data is currently being investigated in a separate paper.

\section{Elemental Abundances}
\begin{figure*}[htbp]
\centering
\includegraphics[trim=0 0 0 0, clip, width=1.1\textwidth]{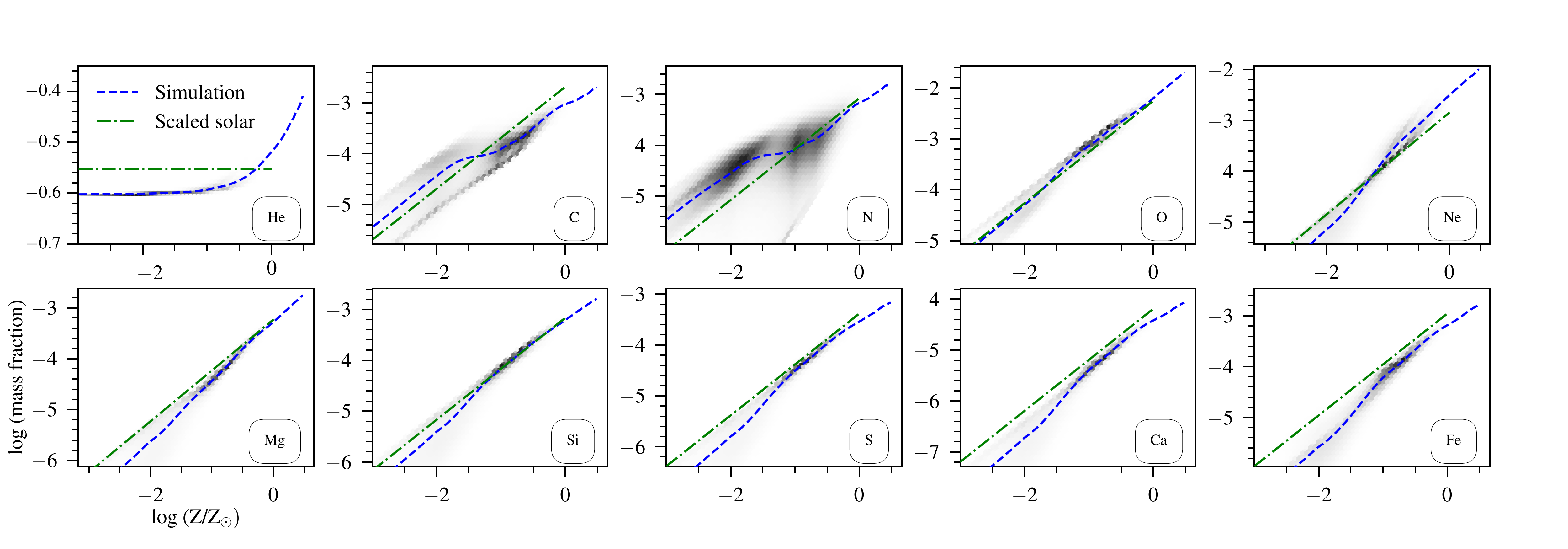}
\caption{Gray hexbins represent the mass fractions of each elements tracked in \simba as function of metallicity for all gas fluid elements.
A running mean is determined for each element as a function of metallicity (blue line).
The green lines show the Solar abundance as a function of metallicity, after scaling by the respective factors determined in order to match the abundance ratio
from \simba (blue). These factors yield a mapping function that enables the \ncode{cloudy} input to reflect the abundance ratio of the \simba galaxies that
differ from Solar.
\label{fig:abun}}
\end{figure*}

In \ncode{cloudy}, one can supply the total metallicity, where the abundance of each elements is then scaled assuming the Solar composition.
We account for abundance patterns of galaxies that differ from Solar using the abundance ratio for each elements tracked
in \simba (i.e., He, C, N, O, Ne, Mg, Si, S, Ca, and Fe).
Since the mass fraction of each element varies as a function of metallicity, we determine its running means for each element using all the gas fluid elements of
the \simba galaxies studied here. This yields a function that maps a given metallicity to an abundance pattern (see \Fig{abun}),
such that for a given metallicity of the clouds, the relative elemental abundances in the \ncode{cloudy} input
are scaled to reflect the (average) abundance ratio from \simba.
For more details on this procedure, we refer interested readers to \citet{Olsen17a}.

\clearpage

\section{\Lcii--\,SFR Relation and DTM}
\begin{figure*}[htbp]
\centering
\includegraphics[trim=0 0 0 0, clip, width=.75\textwidth]{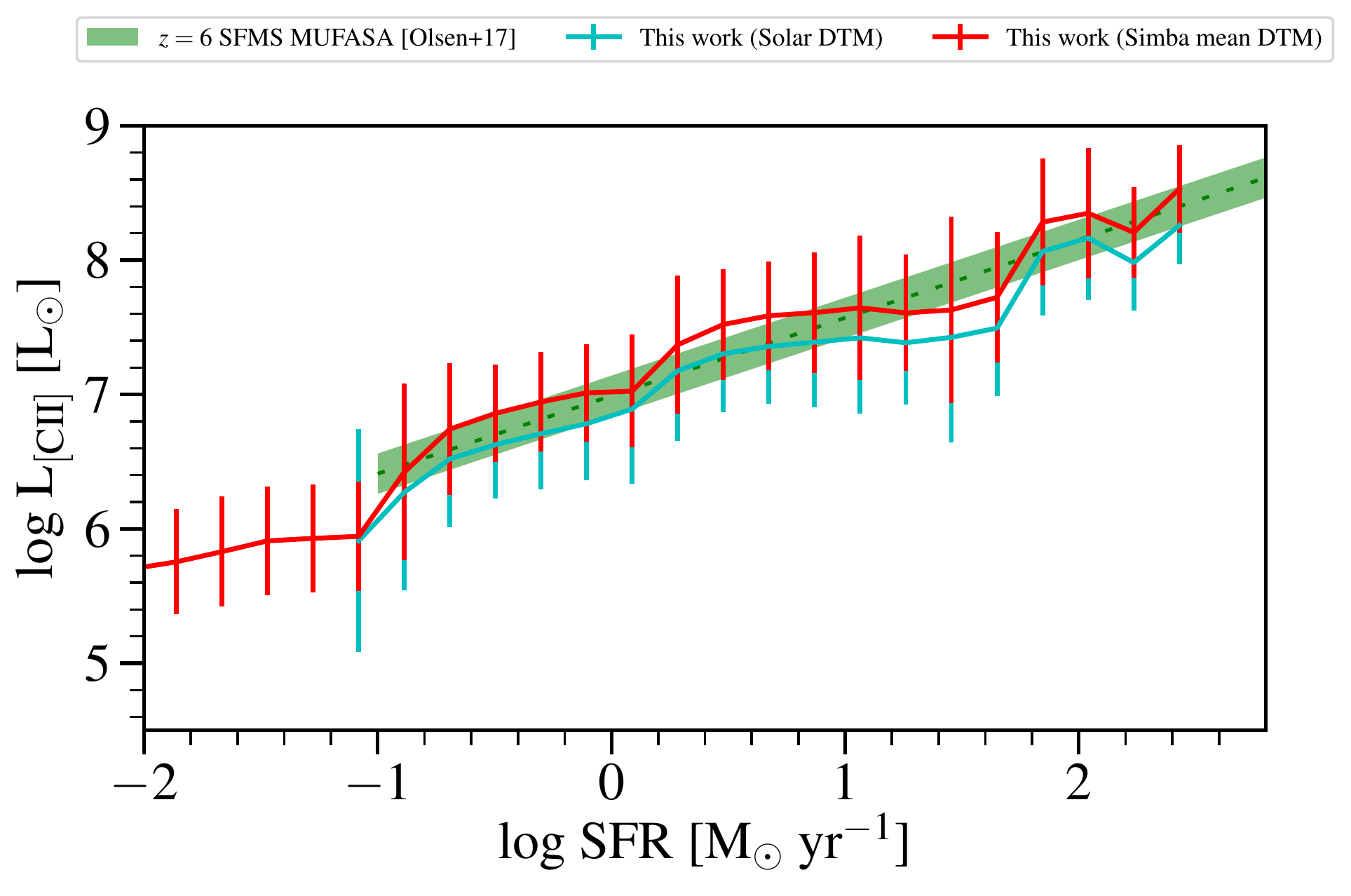}
\caption{\Lcii--\,SFR of \simba galaxies when adopting the mean DTM of the \simba sample (red line) and when adopting
a Solar DTM (cyan line). Adopting the lower DTM results in predicted \cii luminosities at fixed SFR that are about 0.5 lower than those with a Solar DTM.
\label{fig:ciisfr_DTM}}
\end{figure*}

In \Fig{ciisfr_DTM}, we show the effects of adopting a different DTM ratio on the \Lcii--\,SFR relation, in particular,
using a Solar DTM following \citet{Olsen17a} ($\xi_{\rm DTM}$\eq0.46) instead of
the mean DTM ratio found in the \simba sample ($\xi_{\rm DTM}$\eq0.25).
Using the former ratio, the \cii luminosity is approximately $\lesssim$\,0.5\,dex lower
(cf. \citealt{Olsen17a} who finds that the \Lcii is only $\sim$\,0.15\,dex lower at
the same SFR for their \ncode{Mufasa} SFMS sample).

\acknowledgements

We thank Bade Uzgil for providing the relevant numbers from the ASPECS \obs and the referee for providing constructive comments that improved the clarity of this manuscript.
T.K.D.L acknowledges support from the Simons Foundation and the hospitality of the Cosmic DAWN Center and the Danmarks Tekniske Universitet (DTU-Space).
 The Flatiron Institute is supported by the Simons Foundation.
T.K.D.L, K.P.O, and T.R.G
thank NORDITA for its hospitality during the NORDITA 2019 program “Zoom-in and Out: from the Interstellar Medium to the Large Scale Structure of the Universe”.
This research has made use of NASA's Astrophysics Data System Bibliographic Services.
We acknowledge use of the Python programming language \citep{VanRossum1991}, Astropy \citep{astropy},
Matplotlib \citep{Hunter2007}, NumPy \citep{VanDerWalt2011}, SciPy \citep{scipyref}, and \ncode{yt} \citep{Smith09a,Turk11a}.
We thank Robert Thompson for developing the Python package \ncode{caesar}.

\bibliographystyle{aasjournal}
\bibliography{bibs}

\end{document}

%% file: preamble.tex
\usepackage[breaklinks,colorlinks,citecolor=blue,linkcolor=red]{hyperref} 
\usepackage{etoolbox}

\makeatletter

\patchcmd{\NAT@citex}
  {\@citea\NAT@hyper@{%
     \NAT@nmfmt{\NAT@nm}%
     \hyper@natlinkbreak{\NAT@aysep\NAT@spacechar}{\@citeb\@extra@b@citeb}%
     \NAT@date}}
  {\@citea\NAT@nmfmt{\NAT@nm}%
   \NAT@aysep\NAT@spacechar\NAT@hyper@{\NAT@date}}{}{}

\patchcmd{\NAT@citex}
  {\@citea\NAT@hyper@{%
     \NAT@nmfmt{\NAT@nm}%
     \hyper@natlinkbreak{\NAT@spacechar\NAT@@open\if*#1*\else#1\NAT@spacechar\fi}%
       {\@citeb\@extra@b@citeb}%
     \NAT@date}}
  {\@citea\NAT@nmfmt{\NAT@nm}%
   \NAT@spacechar\NAT@@open\if*#1*\else#1\NAT@spacechar\fi\NAT@hyper@{\NAT@date}}
  {}{}

\makeatother
\usepackage{tabu}
\usepackage{tabularx}
\usepackage{booktabs}
\usepackage{xspace}
\usepackage{apjfonts}
\usepackage{amssymb, amsmath} 
\usepackage{natbibspacing, natbib}
\usepackage{aas_macros} 
\usepackage{graphics,graphicx}
\newcommand{\cmpc}{\ensuremath{\mathrm{cMpc}\,\mathrm{h}^{-1}}}
\newcommand{\pmOne}{\mbox{$^{-1}$}\xspace}
\newcommand{\rarr}{$\rightarrow$}
\newcommand{\sig}{$\sigma$}

\newcommand{\aco}{\mbox{CO\,($J$\,=\,1\,\rarr\,0) }}

\newcommand{\cii}{[C{\scriptsize II}]\xspace}

\def\cc{${\rm cm}^{-3}$\xspace}

\newcommand{\mgas}{\mbox{$M_{\rm gas}$}\xspace}

\newcommand{\mstar}{\mbox{$M_*$}\xspace}
\newcommand{\Msun}{\mbox{$M_{\odot}$}\xspace}

\newcommand{\Lsun}{\mbox{$L_{\odot}$}\xspace}

\newcommand{\kms}{km\,s$^{-1}$\xspace}
\newcommand{\bmm}{beam$^{-1}$\xspace}

\newcommand{\Lcii}{\mbox{$L_{\rm [CII]}$}\xspace}

\newcommand{\E}[1]{$\times10^{#1}$}

\newcommand{\eq}{\,=\,}
\newcommand{\ssim}{\,$\sim$\,}

\newcommand{\Fig}[1]{Figure~\ref{fig:#1}}

\newcommand{\Tab}[1]{Table~\ref{tab:#1}}
\newcommand{\Sec}[1]{\S\ref{sec:#1}}
\newcommand\tna{\,\tablenotemark{a}}

\newcommand{\ncode}[1]{{\sc #1}}

\newcommand{\z}{$z$\xspace}

\newcommand{\obs}{observations\xspace}

\newcommand{\athighz}{at high redshift\xspace}

\newcommand{\SF}{star formation\xspace}

\newcommand{\simba}{\ncode{simba}\xspace}
\newcommand{\mufasa}{\ncode{mufasa}\xspace}
\newcommand{\sigame}{\ncode{s\'{i}game}\xspace}

%% file: Table_paramSpace.tex
\def\arraystretch{1.2}
\begin{deluxetable}{ccc}
\centering
\tabletypesize{\scriptsize}
\tablewidth{0.5\textwidth}
\tablecolumns{3}
\tablecaption{Parameter space probed by our \simba galaxy sample at $z$\ssim6.}
\tablehead{
\colhead{Properties} &
\colhead{Units} &
\colhead{Ranges}
}
\startdata
$\log M_{\rm halo}$     & \Msun                       & [9.02 12.36]  \\  [0.3em]
SFR$_{\rm 10}$     & \Msun\,yr\pmOne             & [0.01, 708]    \\  [0.3em]
SFR$_{\rm 100}$    & \Msun\,yr\pmOne             & [0.007, 329]    \\  [0.3em]
$\log \Sigma_{\rm SFR}$ & \Msun\,yr\pmOne\,kpc$^{-2}$ & [-2.60, 1.31]    \\  [0.3em]        
$\log$ \mstar             & \Msun                       & [7.18, 10.72]  \\  [0.3em]
$\log$ \mgas       & \Msun                       & [7.27, 10.46]  \\  [0.3em]
$\langle Z_{\rm gas}\rangle_{\rm SFR}$              &  $Z_\odot$                                 & [0.06, 1.86]   \\ 
%
\enddata
\label{tab:param}
\end{deluxetable}

%% file: Table_models.tex
\def\arraystretch{1.2}
\begin{deluxetable*}{ccccccc}
\centering
\tabletypesize{\scriptsize}
\tablecolumns{7}
\tablecaption{Quick comparison of approaches between existing models in simulating \cii emission at $z$\ssim6.}
\tablehead{
\colhead{} &
\colhead{Vallini15} &
\colhead{Olsen17} &
\colhead{Pallottini17a,b}  &
\colhead{Lagache18}  &
\colhead{Popping19} &
\colhead{This work}
}
\startdata
Simulation technique  & SPH & MFM   &    AMR &    SAM    & SAM    &  MFM  \\  [0.3em]
Gas mass/cell resolution (\Msun) & 10$^{5.1}$ &  10$^{5.3}$ & 10$^{4.1}$ & --- & ---  &  10$^{5.5-7.3}$ \\  [0.3em]
Min. gravitational softening length ($h$\pmOne~kpc) & ---  & 0.01 &  ---  & --- & --- & 0.125--0.5\\  [0.3em]
Stellar emission calculation  &   \ncode{Licorice}, \ncode{starburst99}  &  \ncode{starburst99} & \ncode{starburst99} & \ncode{BC03}\tna & scaled by SFR volumetric density & \ncode{starburst99} \\  [0.3em]
Line emissivity calculation & \ncode{ucl\_pdr} & \ncode{cloudy} & \ncode{cloudy} & \ncode{cloudy} (PDR only) & \ncode{despotic} & \ncode{cloudy}\\  [0.3em]
Number of Galaxies    & 1   & 30      & 2    & 25,000 & $>$230,000 & 11,137 \\
\enddata
\label{tab:model}
\tablecomments{Literature models from \citet{Vallini15a, Olsen17a, Pallottini17a, Pallottini17b, Lagache18a, Popping19a}.}
\tablenotetext{a}{\citet{BC03a}.}
\end{deluxetable*}